\documentclass{article}

\usepackage[preprint]{neurips_2026}

\usepackage[utf8]{inputenc} 
\usepackage[T1]{fontenc}    
\usepackage{hyperref}       
\usepackage{url}            
\usepackage{booktabs}       
\usepackage{amsfonts}       
\usepackage{nicefrac}       
\usepackage{microtype}      
\usepackage{xcolor} 

\usepackage{amsmath, amssymb, amsthm, bm}
\usepackage{cleveref}
\usepackage{thm-restate}
\usepackage{graphicx}
\usepackage{enumitem, comment}
\usepackage{subcaption}

\hypersetup{
    colorlinks,
    citecolor=blue,
    linkcolor=violet,
    urlcolor=blue,
    hypertexnames=false,
    bookmarksnumbered=true,
    pdfpagemode=UseNone,
    pdfstartview=FitW,
}

\theoremstyle{plain}
\newtheorem{theorem}{Theorem}
\newtheorem{lemma}{Lemma}
\newtheorem{definition}{Definition}
\newtheorem{assumption}{Assumption}

\newtheorem{claim}[theorem]{Claim}

\newtheorem{example}{Example}

\usepackage[ruled,noend,linesnumbered]{algorithm2e}
\newcommand*{\abs}[1]{\lvert #1 \rvert}
\newcommand*{\norm}[1]{\lVert #1 \rVert}
\newcommand{\x}{\bm{x}}
\newcommand{\y}{\bm{y}}

\newcommand{\argmax}{\operatornamewithlimits{argmax}}

\newcommand{\yc}[1]{{\color{cyan}[Brian: #1]}}

\allowdisplaybreaks[4]
\newcommand{\OPT}{\operatorname{OPT}}

\newcommand{\dd}{\delta}
\newcommand{\DD}{\Delta}

\title{The Stability of Data Exchange in\\ Competitive Markets}

\author{%
  Yuanchen Brian~Tang\\
  University of Illinois Urbana-Champaign\\
  \texttt{yctang1@illinois.edu} \\
    \And
  Jiaxin~Song\\
  University of Illinois Urbana-Champaign\\
  \texttt{jiaxins8@illinois.edu} \\
  \And Bhaskar Ray~Chaudhury\\
  University of Illinois Urbana-Champaign\\
  \texttt{braycha@illinois.edu} \\
  \And Ruta~Mehta\\
  University of Illinois Urbana-Champaign\\
  \texttt{rutameht@illinois.edu} \\
}

\begin{document}

\maketitle

\begin{abstract}
We study the stability of data exchange among competing firms. Unlike tangible goods, data is non-rival: sharing it does not deprive the giver of its use. On one hand, competing firms-- often operating as machine-learning–driven predictors that collect observations from a shared environment to solve similar tasks-- can improve their predictive accuracies from pooling data. On the other hand, they compete for market share, which may depend on their relative predictive performance. This raises the central question of this paper: {\em can there be a stable data exchange 
in such { competitive} environments?}

We formalize this via a model of $n$ firms exchanging fractions of their datasets, with predictive performance feeding into market competition. We adopt core stability as our solution concept, accounting for externalities via worst-case responses by non-participating firms. Our main results reveal a sharp contrast driven by demand structure: (i) in \emph{fully rational} markets, where customers flock entirely to the top-performing firm, a core-stable exchange always exists for any monotone continuous prediction functions; (ii) in \emph{partially rational} markets with logit-style demand, core-stable exchanges may fail to exist—and even approximate cores can be empty. Our impossibility results suggest that internal profit-sharing within coalitions can restore stability in partially rational markets. We formalize this intuition by showing that when co-operating coalitions equally redistribute market-share revenues among members, a core-stable data exchange exists. Finally, we propose a natural exchange dynamic and demonstrate empirically that it converges to a core-stable data exchange in fully rational markets on simulated instances derived from real datasets.

\end{abstract}

\section{Introduction}
\label{sec:intro}
Data has become a central input in modern production, shaping performance across a wide 
range of industries, including digital platforms, healthcare, transportation, and finance. 
Firms increasingly rely on data-driven models to improve prediction~\cite{brynjolfsson2021power}, 
targeting~\cite{grandhi2021data}, and operational efficiency~\cite{wamba2017big}, making 
access to high-quality data a key determinant of competitive advantage. At the same time, valuable data is often fragmented across firms operating in similar markets, creating strong incentives for data exchange.

\paragraph{Stability of Data Exchange in Competitive Environments.}
This paper investigates the existence and properties of \emph{stable outcomes in data 
exchange among competing firms that operate in similar product markets}. A key feature 
of data is that it is \emph{non-rival}~\cite{jones2020nonrivalry, shapiro1999information}%
\footnote{Consumption by one agent does not affect its availability for another.}: 
pooling datasets does not require firms to forgo or lose access to their own. As a result, 
a data exchange can improve predictive and decision-making performance for all participating 
firms without depleting their individual informational assets~\cite{jones2020nonrivalry}. 
Moreover, the value of data exchange increases with the extent to which firms operate in 
overlapping market spaces: when firms share a market, they serve overlapping customer bases 
and solve essentially the same prediction problems, so data collected by one firm is directly 
relevant for the other's decisions~\cite{gradwohl2023coopetition}. Pooling such data therefore 
increases the effective sample size for a common prediction task, reducing estimation error 
and improving predictive accuracy. \emph{Consequently, the largest technological gains from 
cooperation arise precisely among close competitors with substantial market overlap, rather 
than among firms operating in distinct markets.}

At the same time, improvements in predictive performance need not translate into gains in 
market share, especially when there is a large overlap in the market. While a data exchange 
raises the predictive performance of all firms, firms continue to compete for the same 
customers, so market shares depend on relative-- rather than absolute-- performance.
As a result, even if all firms benefit from data exchange, some may gain less than others 
or even lose market share if competitors are able to leverage the shared data more 
effectively~\cite{gradwohl2023coopetition}. This creates the following fundamental tension: 
\begin{center}
\emph{the largest technological gains arise from data exchange among close competitors, 
yet these gains do not necessarily improve each firm's competitive position. Understanding 
which coalitions and data exchanges remain stable in the face of these incentives is the 
central question explored in this paper.}
\end{center}

We remark that the above phenomenon stands in contrast to the 
exchange of rival tangible goods, where overlap in firms' market spaces typically reduces 
the gains from cooperation. When firms operate in overlapping markets-- for example, 
competing steel producers such as US Steel and ArcelorMittal-- their production 
technologies and input requirements largely coincide. Since physical inputs are rival, 
any exchange involves a transfer of scarce resources: one firm's use comes at the expense 
of the other~\cite{varian1992microeconomic}. As a result, sharing similar inputs primarily 
reallocates existing resources rather than creating new value. Thus, in rival exchange 
environments, gains from trade are driven by differences in endowments or capabilities~\cite{ricardo1817principles, heckscher1919effect}, 
implying that mutually beneficial cooperation is more likely between firms operating in 
non-overlapping markets with complementary needs, rather than among close competitors.

\subsection{Our Contributions}\label{sec:ourContri}
We introduce a stylized model with $n$ firms operating in an overlapping market space, i.e., serving a common customer base. Each firm $i$ performs its prediction task $t_i$ relevant to this shared market. A data exchange is represented by a matrix $\bm{x} \in [0,1]^{n \times n}$, where $x_{i,j}$ denotes the fraction of firm $j$'s data that is shared with firm $i$. Let $\bm{x}_i = (x_{i,1}, x_{i,2}, \dots, x_{i,n})$ denote the data bundle received by firm $i$. Given this data bundle, let $P_i(\bm{x}_i)$ denote the predictive performance of firm $i$, capturing the improvement in accuracy from training on the pooled data. We emphasize that $P_i(\cdot)$ may also depend on firm $i$'s initial data endowment, machine learning model, and available computing resources. We assume that $P_i(\cdot)$ is weakly increasing in each component $x_{i,j}$, reflecting that additional data (weakly) improves prediction. Firms compete for market share based on their predictive performance. Let $u_i(P_1(\bm{x}_1), \dots, P_n(\bm{x}_n))$ denote the market share of firm $i$ after data exchange. We assume that $u_i(\cdot)$ is increasing in own performance $P_i$ and decreasing in rivals' performance $P_j$ for all $j \neq i$, capturing the idea that market outcomes depend on relative predictive advantage.

\paragraph{Coalition.}A coalition is simply a subset of the agents who can choose to coordinate — pooling their actions, resources, or decisions and behaving as a single unit rather than acting independently. We consider coalitions because our agents are self-interested and free to form groups: an outcome that is good in aggregate, or robust to individual deviations, may still be undermined if some subset of agents can break away and do strictly better together. Reasoning about coalitions is what lets us characterize outcomes that are stable against group deviations, not just unilateral ones. This is the standard motivation behind some of the strongest solution concepts in cooperative economics — most notably core stability~\cite{von2007theory,gillies1959solutions}, where an allocation is stable exactly when no coalition can improve on it, and stable matching~\cite{gale1962college,roth1992two}, where stability means no coalition (e.g. a blocking pair) has both the ability and the incentive to deviate. 
\paragraph{Core-Stability.} Within this environment, we adopt core stability---a gold-standard concept from cooperative game theory---as our solution concept. At a high level, a data exchange is core-stable if no coalition of agents has an incentive to deviate by forming a separate, self-contained exchange. In contrast to existing studies on data exchange~\cite{akrami2025theoretical, hakim2026fair, song2025existence}, a key complication arises in our model from externalities: the behavior of agents outside a deviating coalition affects the utilities of those within it. Consequently, when evaluating a potential deviation, agents must take as given the response of those outside the coalition. A natural and credible benchmark is to assume that external agents respond in the most adverse way. This ensures that every agent within the coalition attains a utility level at least as high as the one it anticipates when considering the deviation.

Accordingly, we define a data exchange $\bm{x}$ to be core-stable if there exists no coalition of firms $S \subseteq N$ and alternative exchange $\bm{y}$ (restricted to firms in $S$) such that every firm in $S$ strictly prefers $\bm{y}$ to $\bm{x}$ when firms outside $S$ respond {by inducing the maximum externality on $S$}--
namely, by fully sharing data among themselves.

\paragraph{Market Shares and Customer Rationality.}
We consider a setting in which all firms operate in a common market and compete for the same pool of customers. If customers exhibit \emph{full rationality}, the entire market share is allocated to the firms with the highest predictive accuracy. Formally, given an exchange $\bm{x}$, let $p^*(\bm{x}) = \max_{i} P_i(\bm{x}_i)$, and let $S^* = \{i : P_i(\bm{x}_i) = p^*(\bm{x})\}$ denote the set of top-performing firms. Then,
\[
u_i(P_1(\bm{x}_1), \dots, P_n(\bm{x}_n)) =
\begin{cases}
\frac{1}{|S^*|} & \text{if } i \in S^*, \\
0 & \text{otherwise.}
\end{cases}
\]

We also consider alternate customer behavior, call it the \emph{partially rational} market, where customers respond smoothly to differences in predictive performance. In this case, we model market shares using a logit-style demand function:
\[
u_i(P_1(\bm{x}_1), \dots, P_n(\bm{x}_n)) = \frac{g(P_i(\bm{x}_i))}{\sum_{\ell \in N} g(P_{\ell}(\bm{x}_\ell))},
\]
where $g(\cdot)$ is a strictly increasing function, and the derivative of $g(\cdot)$ capture the degree of customer sensitivity to performance differences.

\paragraph{Our Findings: The Contrast between Full and Partial Rationality.}
We now summarize our main findings on the stability of the \emph{coopetitive} data exchange economy.
\begin{itemize}[leftmargin=*]
    \item As our first main result, we show that a core-stable data exchange always exists in a fully rational market (\Cref{thm:fully_rational_core}), for all monotone and continuous $P_i(\cdot)$. This establishes that stable outcomes for data exchange arise in competitive environments across a broad class of prediction and utility functions. {This is a bit surprising, since the competition is fierce when consumers are fully rational!}

    \item In contrast to this positive result, we show that a core-stable data exchange need not exist in a partially rational market. This highlights that smoother (i.e., more elastic) customer responses can undermine stability. Moreover, this non-existence result holds even when all agents share identical prediction functions\footnote{i.e., agents have identical models and computational resources.} and $g(x) = x^{\rho}$ for any $\rho > 0$. We further strengthen this result by showing that even an approximate core may fail to exist. The primary reason is the following: under partially rational demand, an agent can obtain a strictly positive market share even with a small improvement in predictive performance. This allows highly asymmetric data exchanges to constitute blocking coalitions---where one agent shares a large amount of data while receiving only a small amount in return---still benefiting due to the induced little-performance gain. Such deviations are not viable in fully rational markets, where only the top-performing agents receive any utility. Consequently, partially rational markets admit a richer set of profitable deviations, which ultimately precludes stability (though formalizing this intuition requires non-trivial work; see \Cref{thm:partial_rational}).

    \item The contrast highlights how demand structure shapes cooperation. In fully rational markets, winner-take-all competition limits profitable deviations and supports stable, exclusive data exchange. In partially rational markets, even small performance gains yield positive utility, enabling asymmetric exchanges that destabilize outcomes. This suggests that restoring stability may require coordination mechanisms such as more balanced profit-sharing, which can reduce incentives for such deviations and mimic the equal division of market share among top-performing agents in fully rational markets. We provide a discussion on this in \Cref{sec:partially_rational}-- in particular, we show that when a coalition splits its market share revenue equally among all its participants, then a core-stable data exchange exists.

    \item Finally, we complement our theoretical results with empirical simulations on a concrete machine learning setting. We instantiate our model using the California Housing  dataset~\cite{pace1997sparse}, where $n$ firms each hold a disjoint subset of features and rows, and seek to exchange data to improve a shared regression task. Prediction 
    accuracy is modeled via a Leontief function, capturing the complementary nature of feature-disjoint data. While our proof of existence of a core-stable exchange in the fully rational setting is constructive, it does not directly yield a polynomial-time algorithm (in fact, we show it is unlikely in Theorem~\ref{thm:apxhard}). We therefore propose a natural \emph{peeling algorithm} for computing core-stable outcomes. The algorithm is strikingly efficient in practice: it terminates in under $0.015$ seconds across all tested instance sizes, achieves a core-stability rate of $100\%$ on real-data instances, and $89\%$ on synthetic random instances. This suggests that stable data-exchange outcomes, while potentially hard to compute in the worst case, may be readily achievable in practice via simple, natural dynamics.
  
\end{itemize}

\subsection{Further Related Work}
\label{sec:related_work}

\paragraph{Data exchange market.} 
Our work closely relates to the data exchange models studied in~\cite{akrami2025theoretical,song2025existence,hakim2026fair}, where a data exchange of $n$ agents is also modeled as a matrix $\x\in [0, 1]^{N\times N}$ and each agent obtains an accuracy of $a_i(\x_i)$ as utility.
\cite{akrami2025theoretical} proved the existence of a fair and core-stable data exchange.
\cite{song2025existence} studied the setting when an agent also incurs a cost for data-sharing and showed core-stable data exchange exists under mild conditions.
\cite{hakim2026fair} then introduced a fair-exchange contract where, for every pairwise exchange, each agent receives exactly as many data points as it provides and proved that a pure Nash Equilibrium exists.
One limitation of these works is that they assume no {externalities} ---the data exchange between agents does not affect the utilities of agents outside the exchange.
Our work characterizes the externality by considering a competitive setting in which one's accuracy gain can negatively affect the utility of other agents.

\paragraph{Data economics.} 
There is a growing body of work studying the data economics.
Here, we focus on the works most closely related to our contributions.
In our paper, each agent has a function $P_i$ that measures how much an agent can gain from the shared data.
This connects to a broader literature on the value of data (or information).
Depending on the application context, the value of data has been interpreted as the reduction of uncertainty in prediction~\cite{veldkamp2023valuing, farboodi2023data,farboodi2025valuing,ChaudhuryGMS26}, expected benefits in decision-making~\cite{BabaioffKP12,bergemann2018design}, and competitive advantage relative to peers~\cite{hossain2023equilibrium}.
We refer the readers to \cite{fleckenstein2023review} for a more detailed survey on data valuation.

In addition, with the rapid growth in the demand for data, a substantial body of work~\cite{BabaioffKP12,bergemann2018design,agarwal2019marketplace,hossain2023equilibrium,fallah2024three,chaudhury2026revenue} studied the design of the data market.
A few questions on truthfulness, pricing, and incentives have been studied for data markets.
\cite{AgarwalDS19} introduced a two-sided data marketplace specifically for machine learning data, where the platform collects prediction tasks from data users and addresses these requests by allocating the datasets provided by data sellers.
They also design a truthful mechanism for the platform.
\cite{admati1986monopolistic, admati1990direct,  babaioff2012optimal,bergemann2018design,chaudhury2026revenue} studied the optimal pricing of a monopolist data seller. 
\cite{fallah2024three} considers a setting where data users benefit from platform services in exchange for data and incur privacy loss.


\section{Preliminaries}
\label{sec:model}

{Recall our model from Section \ref{sec:ourContri} with $n$ firms operating in an overlapping market space. A data exchange is represented by $\bm{x} \in [0,1]^{n \times n}$, where $x_{i,j}$ is the fraction of firm $j$'s data that she shares with firm $i$. The data bundle  $\bm{x}_i = (x_{i,1}, x_{i,2}, \dots, x_{i,n})$ received by firm $i$ gives her prediction performance. $P_i(\bm{x}_i)$, where $P_i(\cdot)$ is a non-negative, continuous, non-decreasing function. $P_{-i}(\cdot)$ represents the vector $(P_1(\cdot),\dots,P_n(\cdot))$ excluding $i$-th component. The market share of firm $i$ after the data exchange is $u_i(P_1(\bm{x}_1), \dots, P_n(\bm{x}_n))$, where $u_i(\cdot)$ is increasing in own performance $P_i$ and decreasing in rivals' performance $P_j$ for all $j \neq i$, capturing the competition.}
{We use $\bm 0$ to denote an \emph{null exchange} where $x_{i,j}=0$ for any $i,j$, and $\bm 1$ to denote a \emph{full exchange} where $x_{i,j}=1$ for any $i,j$}

We adopt the concept of \emph{core} from cooperative game theory as our solution concept.

\begin{definition}[Coalition]
    A coalition of a data exchange $\bm{x}$ is defined by $S \subseteq N$ and $\bm{y} = (y_{i,j})_{i,j \in S}$ such that for the induced data exchange $\bar{\y}$ (also denoted by $(\y, \mathbf{1}^{N\setminus S})$) defined by
    $$
    \bar{y}_{i,j} = \begin{cases}
    y_{i,j} & \text{if } i,j \in S \\
    0 & \text{if } i \in S, j \notin S \text{ or } i \notin S, j \in S \\
    1 & \text{if } i,j \notin S
    \end{cases}
    $$
    and for every agent $i \in S$, we have $u_i(P_i(\bar{\y}_{i}), P_{-i}(\bar{\y}_{-i})) > u_i(P_i(\bm{x}_{i}), P_{-i}(\bm{x}_{-i}))$.
\end{definition}

\begin{definition}[Core-stable data exchange]
    A data exchange $\bm{x}$ is \emph{core-stable} if there does not exist any coalition.
\end{definition}

If a data exchange is not core-stable, we refer to the subset of agents $S$ that can deviate as a \emph{blocking coalition}.
We note that core stability actually imposes stricter conditions than a Nash equilibrium since no subset of agents of \emph{any} size has an incentive to deviate.
Moreover, a \emph{null exchange} $\bm{0}$ (where no agent shares any data) may not be a core in many cases. For example, when $n=3$, two agents can form a blocking coalition by actively exchanging with each other and leaving the third player alone.
Also, we relax the definition of core to $\beta$-core when a core does not exist:

\begin{definition}[$\beta$-core-stable data exchange]
A data exchange $\bm{x}$ is \emph{$\beta$-core-stable} if there does not exist a non-empty subset of agents $S \subseteq N$ and an alternative data exchange $\bm{y}$ among agents in $S$ such that $u_i(P_i(\bm{\bar{y}}_{i}), P_{-i}(\bm{\bar{y}}_{-i})) > \beta\cdot u_i(P_i(\bm{x}_{i}), P_{-i}(\bm{x}_{-i}))$ for any $i\in S$.
\end{definition}

\section{Fully Rational Market}
\label{sec:fully_rational_market}
In this section, we study a market where customers are fully rational: they always choose the most accurate predictor, so only top-performing predictors earn positive market share.
If multiple predictors tie for the highest accuracy, the market share is split equally among them.
\begin{definition}[Fully rational market]
    A data exchange market is \emph{fully rational} if customers only choose the predictors with the highest accuracy, so that the agents in $N_{\max} = \mathrm{argmax}_j P_j$ receive positive utilities. 
    We assume the top-performing agents share the market equally and thus receive the same utility $1/\abs{N_{\max}}$.
    Formally, the utility is given by
    $$u_i(P_i, P_{-i}) = \begin{cases}
      1/\abs{N_{\max}}, & \text{if}\ i\in N_{\max},\\
      0, & \text{otherwise.}
    \end{cases}
    $$
\end{definition}

\begin{example}[Three agents]
Consider a data exchange market of $3$ agents, with the prediction accuracy functions defined as $P_i(\bm{x}_i) = \norm{\bm{x}_{i}}_1$.
Consider an initial exchange where agents $1$ and $2$ are fully exchanging their data with each other, i.e., $x_{1,2}=x_{2,1} = 1$ while agent $3$ does not participate in the exchange.
Hence, both the two agents receive a utility of $1/2$.
After agent $3$ joins the market, and the data exchange becomes a full exchange, all three agents have the same highest accuracy, hence, receive the same utility of $1/3$.
\end{example}

Our main result, \Cref{thm:fully_rational_core}, shows that a core always exists in a fully rational market. {This result applies to arbitrary continuous monotone predictor functions; in particular, it does not require linearity, concavity, differentiability, or symmetry.}

\begin{restatable}{theorem}{FullyRationalCore}
    \label{thm:fully_rational_core}
A core always exists in a fully rational market. 
\end{restatable}

The existence of a core-stable exchange hinges on a hierarchical structure latent in the market. Intuitively, one can rank coalitions by size, with smaller coalitions occupying higher levels of the hierarchy. A blocking deviation corresponds to a coalition moving \emph{up} this hierarchy---forming a smaller, more exclusive group that all members strictly prefer. The key observation is that such upgrades are \emph{irreversible}: once a coalition moves to a higher level, no further profitable deviation can pull it back down. This monotonicity ensures that the deviation process terminates, and the coalition at the top of the hierarchy---the one no subset of agents wishes to abandon---constitutes a core-stable exchange.
\looseness=-1

In the remainder of this section, we give an intuitive overview of the main ideas and concepts required to prove \Cref{thm:fully_rational_core}.
The full proof is deferred to Appendix~\ref{app:sec3}. Our proof is based on a constructive approach, where we first identify a set of \emph{coalition candidates} that can potentially form blocking coalitions and then show that a core can be constructed by selecting an appropriate coalition candidate and achieving the optimal data exchange within it.

\paragraph{Coalition candidates.}  
To identify potential blocking coalitions, we observe that any blocking coalition $S$ must satisfy two conditions: (i) all agents in $S$ achieve the same highest accuracy, and (ii) agents outside $S$, who fully exchange among themselves, cannot have higher accuracy.
This motivates the following domain: for a subset $S$ and a target accuracy $v$, define 
\[
\Delta^{S,v} = \left\{\,
\x^S \in [0,1]^{|S|\times|S|}
:
\begin{array}{ll}
P_i(\x_i^S,\mathbf{0}^{N\setminus S}) = v, & \forall i\in S,\\[4pt]
P_j(\mathbf{0}^{S},\mathbf{1}^{N\setminus S}) \le v, & \forall j\notin S
\end{array}
\right\}
\]

Intuitively, $\Delta^{S,v}$ captures all internal exchanges where every agent in $S$ reaches the same accuracy $v$ and no agent outside $S$ can exceed this accuracy by fully exchanging with each other.
We say $S$ is a \emph{coalition candidate} if $\Delta^{S, v}$ is non-empty for some $v\ge 0$ and denote their collection by $\mathcal{C}$.

\begin{lemma}
\label{lem:coalition_candidate}
If $S$ forms a blocking coalition, then $S\in\mathcal{C}$.
\end{lemma}
\begin{proof}
Suppose $S$ forms a blocking coalition by deviating to $\y^S$.
For $S$ to block, every agent in $S$ must reach the same top accuracy $v$ (otherwise, some member does not benefit).
This means $\y^S \in \Delta^{S,v}$, so $S\in\mathcal{C}$ and $\y^S$ is feasible for \ref{prog:1}.
\end{proof}

{
If \(\mathcal C=\emptyset\), then by \Cref{lem:coalition_candidate} no blocking coalition can exist. Hence every data exchange is core-stable, and the theorem follows immediately. Therefore, for the remainder of this section, we assume \(\mathcal C\neq\emptyset\).}

\paragraph{Maximum accuracy among coalition candidates.}
For each coalition candidate $S\in \mathcal{C}$, we identify the internal exchange that maximizes the accuracy all members of $S$ can simultaneously achieve:
\begin{align}
\max_{\x^S,\, v}\ v \quad \text{subject to } \x^S \in \Delta^{S, v} \tag{Program (1)} \label{prog:1}
\end{align}
Let $\x^S_{\text{opt}}$ denote the optimal solution and $P_S$ the optimal value. 
Let $u_S$ be the corresponding utility each member of $S$ obtains under the data exchange $(\x^S_{\text{opt}}, \mathbf{1}^{N\setminus S})$.
Intuitively, $u_S$ measures how well $S$ can do in the best case if they deviate.
We call $S$ a \emph{strong coalition} if $u_S = 1/\abs{S}$ (i.e., members of $S$ become the sole top performers), and $S$ is a \emph{weak coalition} if $u_S < 1/\abs{S}$.
The following lemma justifies why $P_S$ and $u_S$ are sufficient to characterize all possible deviations: any blocking coalition must be a candidate in $\mathcal{C}$, and its best-case outcome is already captured by $\x_{\mathrm{opt}}^S$.

\begin{restatable}{lemma}{lemDominate}
\label{lem:x_opt_dominate}
If $S$ forms a blocking coalition by deviating to $\y^S$, then for every agent $i\in S$,
\[
P_S \ge P_i(\y_i^S,\,\mathbf{0}^{N\setminus S})
\qquad\text{and}\qquad
u_S \ge u_i(\y^S,\,\mathbf{1}^{N\setminus S}).
\]
\end{restatable}

\begin{proof}[Proof sketch.]
By \Cref{lem:coalition_candidate}, $S\in\mathcal{C}$ and $\y^S$ is feasible for \ref{prog:1}.
The first inequality then follows from the optimality of $P_S$.
For the second, agents outside $S$ share no data with $S$ and fully exchange among themselves, so their accuracy is the same under $(\x_{\mathrm{opt}}^S, \mathbf{1}^{N\setminus S})$ and $(\y^S, \mathbf{1}^{N\setminus S})$.
Since agents in $S$ achieve weakly higher accuracy under $\x_{\mathrm{opt}}^S$, their rank in the market is at least as high, giving $u_S \ge u_i(\y^S, \mathbf{1}^{N\setminus S})$.
\end{proof}

With \Cref{lem:x_opt_dominate}, checking whether a blocking coalition exists reduces to comparing $u_S$ across candidates in $\mathcal{C}$.
Building on this, we show how to construct a core from the candidate set $\mathcal{C}$.

\paragraph{Construction of core.}
The idea is to select a candidate with the highest utility $u_S$ and the largest size $\abs{S}$ among those with the same highest utility.
Particularly, consider the following candidate:
\begin{align}
S^* = \argmax_{\abs{S}} \left\{S\in \argmax_{S\in \mathcal{C}} u_S \right\}\,.\tag{Program (2)} \label{prog:2}
\end{align}
We show that the data exchange $(\x_{\text{opt}}^{S^*}, \mathbf{1}^{N\setminus S^*})$ induced by $S^*$ forms a core in the following lemma.
The proof is deferred to Appendix~\ref{app:sec3}.
\begin{restatable}{lemma}{lemCore}
\label{lem:core}
The data exchange $(\x_{\text{opt}}^{S^*}, \mathbf{1}^{N\setminus S^*})$ is a core for a fully rational market.    
\end{restatable}

\noindent \emph{Proof sketch.}
Suppose for contradiction that some coalition $S$ blocks $(\x_{\text{opt}}^{S^*}, \mathbf{1}^{N\setminus S^*})$.
One can verify that any blocking coalition must lie in $\mathcal{C}$, so \Cref{lem:x_opt_dominate} applies.
We consider three cases.

\textbf{Case I ($S\cap S^*\neq \emptyset$).}
Any agent $i\in S\cap S^*$ strictly improves from $u_{S^*}$, so \Cref{lem:x_opt_dominate} gives $u_S \ge u_i > u_{S^*}$, contradicting the maximality of $u_{S^*}$.

\textbf{Case II ($S\cap S^*=\emptyset$, $S$ strong).}
A strong coalition strictly beats all outsiders. But $S$ and $S^*$ are disjoint, so each lies in the other's outsiders, giving $P_{S^*} \ge P_S > P_{S^*}$, a contradiction.

\textbf{Case III ($S\cap S^*=\emptyset$, $S$ weak).}
{The remaining scenario involves a tie.}
The same chain now holds with equality: $P_{S^*}=P_S=:\mu$.
Since agents in $S$ and $S^*$ each achieve $\mu$ independently, every member of $S\cup S^*$ can achieve accuracy at least $\mu$, in the optimal exchange $x_{\text{opt}}^{S\cup S^*}$.
Agents outside $S\cup S^*$ have fewer data sources than agents outside $S^*$ alone, so by monotonicity their accuracy is at most $\mu$. $u_{S\cup S^*}\ge u_{S^*}$ follows that no non-top outsiders will become top under the new exchange.
However, $\lvert S\cup S^*\rvert>\lvert S^*\rvert$, contradicting the size-maximality of $S^*$.
\hfill$\square$

\section{Partially Rational Market}
\label{sec:partially_rational}

In practice, customers do not always choose the most accurate predictor---their decisions can also be influenced by factors such as user experience, brand reputation, etc.
As a result, predictors with suboptimal accuracy may still attract positive market share, in contrast to the fully rational market studied in the previous section.
In this section, we consider such a \emph{partially rational market}.

\begin{definition}[Partially rational market]
A data exchange market is \emph{partially rational} if the customers may also choose predictors with suboptimal accuracy.
Particularly, the utility function $u_i$ is monotonically increasing in $P_i$ (higher accuracy yields more utility), and agents with suboptimal accuracy can still receive positive utility, i.e., $u_i(P_i, P_{-i}) > 0$ is possible even when $P_i < \max_{j} P_j$.
\end{definition}

We first give some high-level intuition on why achieving a core-stable data exchange is more challenging in a partially rational market.
In a fully rational market, only agents with the highest accuracy obtain positive utility, so a deviating coalition must significantly restructure the data exchange to alter accuracy rankings and gain utility.
In contrast, in a partially rational market, even agents with suboptimal accuracy receive positive utility, making the market more fragile---even a minor change in the exchange can affect utilities and create incentives for deviation.
Particularly, a core under a fully rational market may break down under a partially rational market.

\begin{example}
    Consider a symmetric market of three agents, where $P_i(\x_i) = \norm{\x_i}_1$.
    By the analysis of \Cref{sec:fully_rational_market}, a core can be achieved when agents 1 and 2 fully exchange their data with each other, leaving agent 3 out.
    Hence, both agents 1 and 2 receive a utility of $1/2$.
    Agent 3 cannot form a deviation with either of them while simultaneously ensuring its own utility is positive, and the other agent's utility strictly increases.

    However, in a partially rational market, the above argument no longer holds. Agent 3 can easily incentivize Agent 1 to deviate by sharing its entire dataset with Agent 1 while requesting only a small amount of data in return. As a result, Agent 1 achieves the highest accuracy strictly and thus receives a utility greater than $1/2$, while Agent 3 also obtains a positive utility.
\end{example}

The above example suggests that a core may not always exist in a symmetric partially rational market.
Our main result for this section confirms this and further characterizes the best approximation guarantee achievable.
Precisely, we focus on the following natural family of utility functions, where each agent's market share is proportional to a transformation of its accuracy:
\begin{align}\label{eq:prop_utility}
u_i(\bm{P}) = \frac{g(P_i)}{\sum_{j\in N}g(P_j)},    
\end{align}
where $g(\cdot)$ is a strictly increasing continuous function with $g(0) = 0$.
If \(\sum_{j\in N}g(P_j)=0\), we define \(u_i(\bm P)=1/n\) for \(i\in N\).
We show that while a core may not exist, a $\beta$-approximate core is always achievable via full exchange, and we give a tight characterization of the minimum $\beta$, as follows.

\begin{restatable}{theorem}{thmPartialRational}
\label{thm:partial_rational}
In a {partially rational market, assuming \(n\ge 3\), and
\(P_i(\x_i)=\sum_{j\neq i}x_{i,j}\) for all \(i\). 
Then a \(\beta\)-core exists if and only if
$
\beta\ge
\beta^*
:=
\max_{1\le k\le n}
\frac{n g(k-1)}
{A_k+k g(k-1)},
$
where
$
A_k=
\begin{cases}
(n-k)g(n-k-1), & k<n,\\
0, & k=n.
\end{cases}
$
Moreover, this threshold is tight for 
a full exchange.}

\end{restatable}


{The above threshold is strictly larger than \(1\), since the numerator is larger than the denominator when $\frac n2<k<n$. Hence an exact core need not exist. Also, the above bound is tight for the particular form of $P_i$. More generally, for any symmetric monotone continuous predictor, the analysis depends only on the composed values
\(
G_m := g(P_i(\underbrace{1,\dots,1}_{m},\underbrace{0,\dots,0}_{n-1-m})),
\)
that is, the transformed market weight of an agent who receives data from \(m\) other agents. Thus the same characterization extends by replacing \(g(m)\) with \(G_m\).
A detailed discussion is deferred to Appendix~\ref{app:sec4}}.



Now we provide a proof sketch for \Cref{thm:partial_rational}.
First, we define the incentive ratio of a coalition $S$ deviating from exchange $\x$ as the minimum utility improvement factor across members of $S$:
\begin{align*}
\rho_S(\x) = \sup_{\y^S} \min_{i\in S} \frac{u_i(\y^S, \mathbf{1}^{N\setminus S})}{u_i(\x)} \tag{incentive ratio}\,.    
\end{align*}

Intuitively, this measures how much each member of $S$ improves. 
A $\beta$-blocking coalition is then a coalition with an incentive ratio of at least $\beta$.
For a data exchange $\x$ and coalition size $k$, let $\beta_{k,\mathbf{x}}^*$ denote the supremum incentive ratio achievable by any $k$-sized coalition deviating from $\x$:  $\beta_{k, \mathbf{x}}^* = \sup_{S:\,|S|=k} \rho_S(\x)$.
Then the minimum $\beta$ for which a $\beta$-core exists is   
\[
\beta^* = \inf_{\x}\, \max_{k}\, \beta_{k,\x}^*.
\]

Next, we show how to calculate $\beta_{k,\x}^*$ for a fixed $\x$ and $k$.
Without loss of generality, assume the accuracies are in a decreasing order: $P_1(\x) \ge \cdots \ge P_n(\x)$, hence, $u_1(\x) \ge \dots \ge u_n(\x)$.
Since agents' accuracy and utility functions are symmetric, it suffices to check the coalition of agents with the lowest utilities, which has the strongest incentive to deviate.
Let $S$ be the set of the $k$ agents with the smallest utilities.
Suppose the agents in $S$ have an incentive ratio of $\beta$ to deviate to a new exchange $\y^S$.
Then we can derive the following constraint $u_i(\y^S, \bm{1}^{N\setminus S}) > \beta u_i(\x)$ for any $i\in S$.
Let $s_j$ be the accuracy of agent $j$ under the new exchange $\y^S$.
We know $s_j = n-k-1$ for any $j\notin S$ since they fully exchange with each other, and $s_j \in [0, k-1]$ for any $j\in S$ since they can only exchange with each other.
Then the constraint can be rewritten as
\begin{align*}
\frac{g(s_i)}{\sum_{j=1}^n g(s_j)} &> \beta \cdot u_i(\x), \quad \forall i\in [n-k+1, n]\\
s_i &\in [0, k-1], \quad \forall i\in [n-k+1, n]
\end{align*}
We fix the sum of $\sum_{j=1}^n g(s_j)$ as $S_g$.
Then the above program can be written as 
\begin{align*}
\mathcal{H} &= \left\{(S_g, g(s_{n-k+1}), \dots, g(s_n)):\beta\cdot u_i(\x)\cdot S_g < g(s_i) \le g(k-1), \forall i\in [n-k+1, n] \right\}\\
\mathcal{P} &=  \{(S_g, g(s_{n-k+1}), \dots, g(s_n)): S_g = (n-k)g(n-k-1) +\sum_{i=n-k+1}^n g(s_i) \}\,.
\end{align*}
Hence, its feasibility is equivalent to checking whether the hyperplane $\mathcal{P}$ passes through the nonempty hypercube $\mathcal{H}$.
Through a delicate calculation, we can derive the maximum value of $\beta$ such that the above program is feasible, which is given by 
$$
\beta_{k,\x}^* = \frac{g(k-1)}{ (n-k)g(n-k-1)u_{n-k+1}(\x) + g(k-1)  \sum_{i\in [n-k+1:n]}u_i(\bm x)}.
$$

Notice that $\beta_{k, \x}^* \ge \beta_{k, \bm{1}}^*$ holds for any data exchange $\x$ and $k>\frac n2$. Additionally, $\beta_{k,\bm 1}^*\le 1$ for $k\le \frac n2$.
Thus, we have $\max_{k} \beta^*_{k, \x} \ge \max_{k>\frac n2} \beta^*_{k, \x} \ge \max_{k>\frac n2} \beta^*_{k, \bm{1}} = \max_{k} \beta^*_{k, \bm{1}}.$ Therefore, we have
$\beta^* = \max_{k} \beta^*_{k, \bm{1}}.$
We can conclude that a $\beta$-core exists if and only if $\beta \ge \max_{k} \beta_{k,\bm 1}^*$, and the equality is achieved when $\x$ is a full exchange.

\paragraph{Profit distributions inside coalitions can restore stability.} Although core stability need not exist in partially rational markets, natural mechanisms can restore it. The fundamental obstruction is that partially rational markets admit highly asymmetric coalitional deviations: an agent with zero utility may be willing to share all of their data with another agent in exchange for only a small fraction of data in return, since even a marginal improvement in predictive performance yields strictly positive market share. Such deviations are unattractive in fully rational markets, where only top-performing agents receive any utility, but are viable here precisely because demand responds smoothly to performance differences.
A natural remedy is to require that any deviating coalition splits its aggregate utility \emph{equally} among its members. Under equal profit-sharing, the asymmetric deviations described above become unappealing: an agent already enjoying high utility would receive less under equal division than under the current arrangement, removing their incentive to participate in such a coalition. We formalize this intuition in the following theorem (full details and proof are deferred to~\Cref{app:equ})

\begin{theorem}[informal]\label{thm:equ}
    The partitioned equalized (partially rational) market admits a non-empty core assuming equalized coalitions.
\end{theorem}

\section{Empirics}
\label{sec:empirics}

\vspace{-0.2cm}

{We note that while solving \ref{prog:2} seems to be the natural route to arrive at a core, we show in \Cref{app:hard}, that this is unlikely-- Solving \ref{prog:2} is APX-hard.} Therefore, in this section, we propose a natural and simple heuristic algorithm, called the \emph{peeling algorithm}, to compute a core, and validate its empirical performance on a data-exchange market constructed using a real-world dataset. 
The idea of the peeling algorithm is to start with a coalition and keep eliminating the one with the least accuracy until it becomes core-stable.
We aim to answer the following questions: \textbf{Q1)} How does the peeling algorithm perform in practice, in terms of the running time and the probability of finding a core; \textbf{Q2)} How likely is a full exchange of a core in a partially rational market?

\noindent \textbf{Setup.} We simulate a data exchange market from the California Housing dataset~\cite{pace1997sparse}, which contains 20,640 observations on 8 features and one target variable (median house value).
We partition the agents into $T = 3$ feature groups: (G1): \texttt{MedInc}, \texttt{HouseAge}, \texttt{AveRooms}; (G2): \texttt{AveBedrms}, \texttt{Population}, \texttt{AveOccup}; (G3): \texttt{Latitude}, \texttt{Longitude}.
Each group $t$ owns a distinct subset of feature columns $\mathcal{F}_t$; the four agents within a group each own a disjoint partition of the 20,640 rows, so every agent holds a local dataset $\mathcal{D}_i = \{(\bm{x}^{(\mathcal{F}_{t_i})}_k, y_k)\}$ consisting of their rows restricted to their group's features.
Each agent's goal is to train a regression model predicting median house value from all $8$ features, which requires collecting records across all three feature groups via data exchange.
Thus, we simulate the prediction accuracy via a Leontief form: $P_i(\x_i) = \min_{t\in [T]}(\alpha_{t, i}\sum_{j\in G_t} x_{i,j})$. 
To estimate $\alpha_{t, i}$, we consider the average marginal contribution of one data record from group $t$, assuming that all other feature groups fully share their data with agent $i$.
All experiments are conducted on a MacBook Pro with an Apple M3 chip and 18GB unified memory.

\noindent \textbf{The peeling algorithm.} 
For computing the core in a fully rational market, 
we propose the peeling algorithm (the pseudocode is put in Algorithm~\ref{alg:peeling}, Appendix~\ref{sec:pseudo_code_peeling}) as follows:
At a high level, instead of identifying $\max_T P_T$, it aims to identify a small subset $S$ such that $P_S>\max_{T\subseteq N\setminus S}P_T$. 
In fact, such a subset with minimum size can be similarly proven to satisfy the core condition. 
Although this problem appears to be computationally hard, the idea of the peeling algorithm is to approximate this optimality by only focusing on candidates along the peeling path.
Particularly, it initializes with the whole set $N$ and keeps track of an answer subset $S$ for a potential solution. In each iteration, the dynamics update $S$ by eliminating the bottleneck agent of $P_S$, i.e., the agent with the worst accuracy, if agents in $S$ can beat the best opponent coalition outside $S$.

\begin{figure}[t]
    \centering
    \begin{subfigure}[t]{0.3\textwidth}
        \centering
        \includegraphics[width=\linewidth]{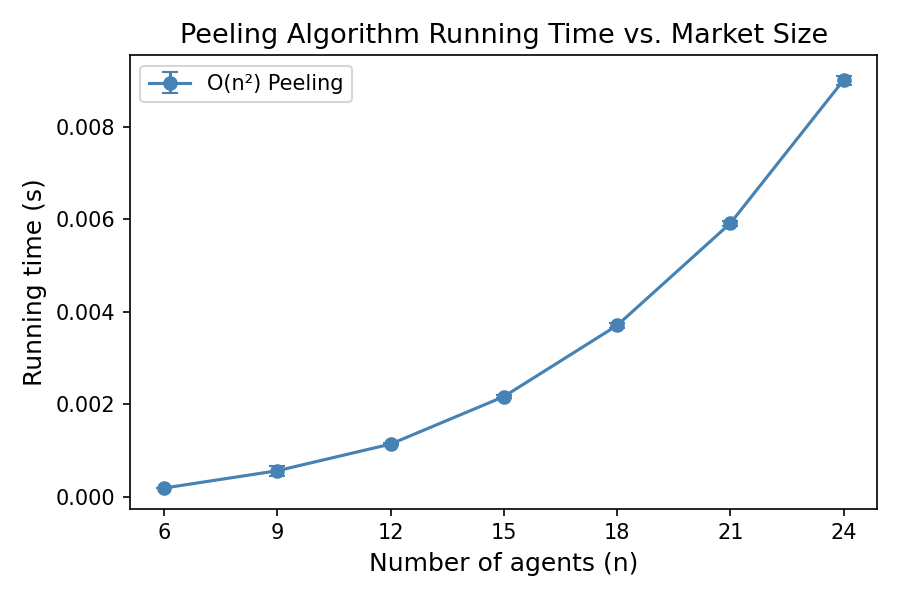}
        \caption{Running time}
        \label{fig:running_time}
    \end{subfigure}
    \hfill
    \begin{subfigure}[t]{0.3\textwidth}
        \centering
        \includegraphics[width=\linewidth]{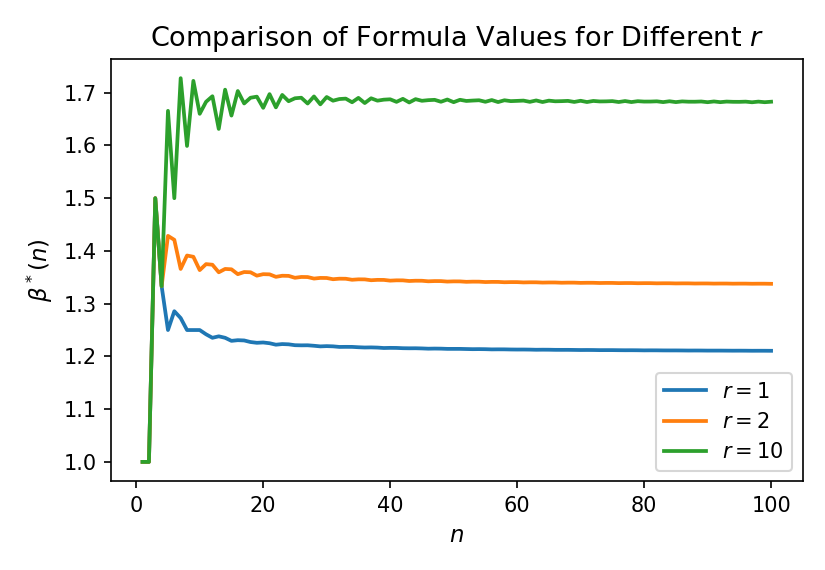}
        \caption{$\beta$ derived from \Cref{thm:partial_rational}}
        \label{fig:symptomatic_analysis}
    \end{subfigure}
    \hfill
    \begin{subfigure}[t]{0.3\textwidth}
        \centering
        \includegraphics[width=\linewidth]{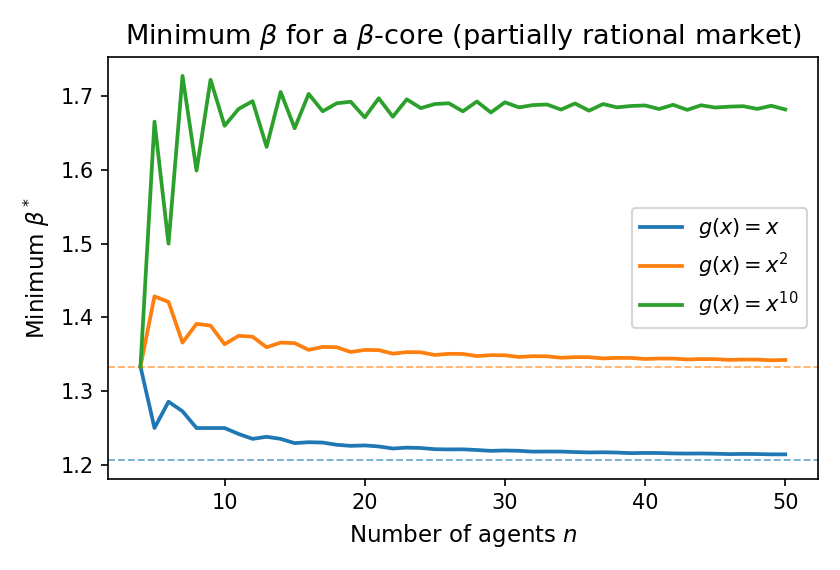}
        \caption{$\beta$ on real data}
    \end{subfigure}
    \caption{Performance of peeling algorithm and Core Approximation in Partially Rational Markets}
    \label{fig:placeholder}
\vspace{-4mm}
\end{figure}

\noindent\textbf{Result analysis.} For \textbf{Q1)}, we run the peeling algorithm for varying numbers of agents (ranging from $n=12$ to $n=30$), with three duplicate experiments per instance. 
The algorithm terminates within $0.015$ seconds across all instances, and the 
running time grows quadratically with $n$, consistent with our theoretical 
analysis (see \Cref{fig:running_time}).

Moreover, we evaluate the probability that the outcome lies in the core.
We fix the number of agents as $12$ and test the success rate on two datasets.
First, we run the peeling algorithm on closed-form accuracy functions derived from the real dataset, where the data is randomly split the data across the agents within each group.
We have tested for $30$ instances and observed that the peeling algorithm found a valid core with a rate of $100\%$, with each output coalition verified to be core-stable.
In addition, to test the robustness of the peeling algorithm on a more general case, we run it on a synthetic dataset, where we use the same number of agents $12$ but generate $\alpha_{t, i}$ uniformly at random from $[0.1, 1]$.
We have run it for $200$ trials, and the rate of the outcome being core-stable is $89\%$.

For \textbf{Q2)}, we investigate the partially rational market proposed in \Cref{sec:partially_rational} with $g(x) = x^r$.
First, we did a symptomatic analysis on the ratio derived from \Cref{thm:partial_rational}.
As plotted in \Cref{fig:symptomatic_analysis}, for the linear case ($r=1$), as $n$ goes to infinity, the threshold converges to $ (\sqrt{2} + 1)/2 \approx 1.207$. Similarly, for $r=2$, the threshold approaches $\beta_{r=2} = 4/3 \approx 1.333$. In general, this asymptotic limit increases monotonically with respect to $r$. These convergences suggest that a constant-factor approximation of the core is consistently achievable.

Meanwhile, we also consider an asymmetric setting in which the accuracy functions $P_i$ are derived from real data. 
We are interested in the empirical approximation ratio achieved by the full exchange mechanism.
Fix the number of agents as $12$ and use binary search to find the smallest $\beta$ with a tolerance of $10^{-4}$ such that $\exists? T, \y. (u_i(\y) > u_i(\x),\forall i\in T)$ is unsatisfiable. 
We observe that, as $r$ goes higher, the value of $\beta$ also goes higher, which aligns our theoretical analysis.

\section*{Acknowledgement}
The research of Bhaskar Ray Chaudhury, Jiaxin Song, and Yuanchen Brian Tang is supported by the NSF CAREER award CCF: 2441580. The research of Ruta Mehta is supported by NSF grant CCF: 2334461.

\bibliographystyle{plain}
\bibliography{reference}

@String{ECold    = {Conf.\ Electronic Commerce (EC)}}

@String{EC       = {Conf.\ Economics and Computation (EC)}}

@String{WWW      = {World Wide Web Conf.\ (WWW)}}

@article{jones2020nonrivalry,
  title={Nonrivalry and the Economics of Data},
  author={Jones, Charles I and Tonetti, Christopher},
  journal={American Economic Review},
  volume={110},
  number={9},
  pages={2819--2858},
  year={2020},
  publisher={American Economic Association}
}

@book{shapiro1999information,
  title={Information Rules: A Strategic Guide to the Network Economy},
  author={Shapiro, Carl and Varian, Hal R},
  year={1999},
  publisher={Harvard Business School Press},
  address={Boston, MA}
}

@article{veldkamp2023valuing,
  title={Valuing Data as an Asset},
  author={Veldkamp, Laura and Chung, Cindy},
  journal={Review of Finance},
  volume={27},
  number={5},
  pages={1545--1562},
  year={2023},
  publisher={Oxford University Press}
}

@article{gradwohl2023coopetition,
  title={Coopetition Against an Amazon},
  author={Gradwohl, Ronen and Tennenholtz, Moshe},
  journal={Journal of Artificial Intelligence Research},
  volume={76},
  pages={735--779},
  year={2023}
}

@book{varian1992microeconomic,
  title={Microeconomic Analysis},
  author={Varian, Hal R},
  edition={3rd},
  year={1992},
  publisher={W. W. Norton \& Company},
  address={New York}
}

@book{ricardo1817principles,
  title={On the Principles of Political Economy and Taxation},
  author={Ricardo, David},
  year={1817},
  publisher={John Murray},
  address={London}
}

@article{heckscher1919effect,
  title={The Effect of Foreign Trade on the Distribution of Income},
  author={Heckscher, Eli F},
  journal={Ekonomisk Tidskrift},
  volume={21},
  pages={497--512},
  year={1919},
  note={Reprinted in: Heckscher, E.F. and Ohlin, B. (1991), Heckscher-Ohlin Trade Theory, MIT Press}
}

@inproceedings{agarwal2019marketplace,
  title={A marketplace for data: An algorithmic solution},
  author={Agarwal, Anish and Dahleh, Munther and Sarkar, Tuhin},
  booktitle={Proceedings of the 2019 ACM Conference on Economics and Computation},
  pages={701--726},
  year={2019}
}

@article{fallah2024three,
  title={On three-layer data markets},
  author={Fallah, Alireza and Jordan, Michael I and Makhdoumi, Ali and Malekian, Azarakhsh},
  journal={arXiv preprint arXiv:2402.09697},
  year={2024}
}

@inproceedings{ChaudhuryGMS26,
    author = {Chaudhury, Bhaskar Ray and Garg, Jugal and Murhekar, Aniket and Song, Jiaxin},
    title = {Data Pricing via Competitive Equilibrium},
    booktitle = {Proceedings of the {ACM} on Web Conference (WWW)},
    year = {2026},
    url = {http://jugal.ise.illinois.edu/papers/www26.pdf}
}

@article{song2025existence,
title={On the Existence and Complexity of Core-Stable Data Exchanges},
author={Song, Jiaxin and Kulkarni, Pooja and Shahkar, Parnian and Chaudhury, Bhaskar Ray},
journal={arXiv preprint arXiv:2509.16450},
year={2025}
}

@inproceedings{akrami2025theoretical,
title={On the Theoretical Foundations of Data Exchange Economies},
author={Akrami, Hannaneh and Chaudhury, Bhaskar Ray and Garg, Jugal and Murhekar, Aniket},
booktitle=EC,
year={2025},
pages={444--444},
doi={10.1145/3736252.3742566}
}

@article{fleckenstein2023review,
title={A review of data valuation approaches and building and scoring a data valuation model},
author={Fleckenstein, Mike and Obaidi, Ali and Tryfona, Nektaria},
journal={Harvard Data Science Review},
year={2023},
volume={5},
number={1},
publisher={The MIT Press},
doi={10.1162/99608f92.c18db966}
}

@article{farboodi2023data,
title={Data and markets},
author={Farboodi, Maryam and Veldkamp, Laura},
journal={Annual Review of Economics},
volume={15},
number={1},
pages={23--40},
year={2023},
publisher={Annual Reviews},
doi={10.1146/annurev-economics-082322-023244}
}

@article{farboodi2025valuing,
title={Valuing financial data},
author={Farboodi, Maryam and Singal, Dhruv and Veldkamp, Laura and Venkateswaran, Venky},
journal={The Review of Financial Studies},
volume={38},
number={3},
pages={938--980},
year={2025},
publisher={Oxford University Press},
doi={10.1093/rfs/hhae034}
}

@article{hossain2023equilibrium,
  title={Equilibrium of data markets with externality},
  author={Hossain, Safwan and Chen, Yiling},
  journal={arXiv preprint arXiv:2302.08012},
  year={2023}
}

@inproceedings{AgarwalDS19,
author = {Anish Agarwal and Munther A. Dahleh and Tuhin Sarkar},
title = {A Marketplace for Data: An Algorithmic Solution},
booktitle = EC,
year = {2019},
pages = {701--726},
publisher = {{ACM}},
doi = {10.1145/3328526.3329589}
}

@article{admati1990direct,
title={Direct and indirect sale of information},
author={Admati, Anat R and Pfleiderer, Paul},
journal={Econometrica: Journal of the Econometric Society},
year={1990},
pages={901--928},
publisher={JSTOR},
doi={10.2307/2938355}
}

@article{admati1986monopolistic,
title={A monopolistic market for information},
author={Admati, Anat R and Pfleiderer, Paul},
journal={Journal of Economic Theory},
volume={39},
number={2},
pages={400--438},
year={1986},
publisher={Elsevier},
doi={10.1016/0022-0531(86)90052-9}
}

@inproceedings{BabaioffKP12,
author = {Moshe Babaioff and Robert Kleinberg and Renato Paes Leme},
title = {Optimal Mechanisms for Selling Information},
booktitle = ECold,
pages = {92–109},
year = {2012},
doi = {10.1145/2229012.2229024}
}

@article{bergemann2018design,
title={The design and price of information},
author={Bergemann, Dirk and Bonatti, Alessandro and Smolin, Alex},
journal={American economic review},
year={2018},
volume={108},
number={1},
pages={1--48},
publisher={American Economic Association},
doi={10.1257/aer.20161079}
}

@article{pace1997sparse,
  title={Sparse spatial autoregressions},
  author={Pace, R Kelley and Barry, Ronald},
  journal={Statistics \& Probability Letters},
  volume={33},
  number={3},
  pages={291--297},
  year={1997},
  publisher={Elsevier}
}

@article{hakim2026fair,
  title={Fair Data-Exchange Mechanisms},
  author={Hakim, Rashida and Papadimitriou, Christos and Yannakakis, Mihalis},
  journal={arXiv preprint arXiv:2602.11417},
  year={2026}
}

@inproceedings{babaioff2012optimal,
  title={Optimal mechanisms for selling information},
  author={Babaioff, Moshe and Kleinberg, Robert and Paes Leme, Renato},
  booktitle={Proceedings of the 13th ACM Conference on Electronic Commerce},
  pages={92--109},
  year={2012}
}

@article{chaudhury2026revenue,
  title={Revenue-Optimal Pricing for Budget-Constrained Buyers in Data Markets},
  author={Chaudhury, Bhaskar Ray and Garg, Jugal and Sharma, Eklavya and Song, Jiaxin},
  journal={arXiv preprint arXiv:2602.13897},
  year={2026}
}

@article{brynjolfsson2021power,
  title={The power of prediction: predictive analytics, workplace complements, and business performance},
  author={Brynjolfsson, Erik and Jin, Wang and McElheran, Kristina},
  journal={Business Economics},
  volume={56},
  number={4},
  pages={217--239},
  year={2021},
  publisher={Springer}
}

@article{wamba2017big,
  title={Big data analytics and firm performance: Effects of dynamic capabilities},
  author={Wamba, Samuel Fosso and Gunasekaran, Angappa and Akter, Shahriar and Ren, Steven Ji-fan and Dubey, Rameshwar and Childe, Stephen J},
  journal={Journal of business research},
  volume={70},
  pages={356--365},
  year={2017},
  publisher={Elsevier}
}

@article{grandhi2021data,
  title={Data-driven marketing for growth and profitability},
  author={Grandhi, Balakrishna and Patwa, Nitin and Saleem, Kashaf},
  journal={EuroMed Journal of Business},
  volume={16},
  number={4},
  pages={381--398},
  year={2021},
  publisher={Emerald Publishing Limited}
}

@incollection{von2007theory,
  title={Theory of games and economic behavior: 60th anniversary commemorative edition},
  author={Von Neumann, John and Morgenstern, Oskar},
  booktitle={Theory of games and economic behavior},
  year={2007},
  publisher={Princeton university press}
}

@article{gillies1959solutions,
  title={Solutions to general non-zero-sum games},
  author={Gillies, Donald B},
  journal={Contributions to the Theory of Games},
  volume={4},
  number={40},
  pages={47--85},
  year={1959}
}

@article{gale1962college,
  title={College admissions and the stability of marriage},
  author={Gale, David and Shapley, Lloyd S},
  journal={The American mathematical monthly},
  volume={69},
  number={1},
  pages={9--15},
  year={1962},
  publisher={Taylor \& Francis}
}

@article{roth1992two,
  title={Two-sided matching},
  author={Roth, Alvin E and Sotomayor, Marilda},
  journal={Handbook of game theory with economic applications},
  volume={1},
  pages={485--541},
  year={1992},
  publisher={Elsevier}
}

\newpage
\appendix
\section{Omitted Proofs in Section 3}
\label{app:sec3}

\lemDominate*
\begin{proof}
Since $(S, \y^S)$ is a blocking coalition, every agent in $S$ receives a higher utility after deviation, which means agents in $S$ must have the same highest accuracy.
Let $v^{(\y)}$ be the common accuracy for agents in $S$ under the data exchange $\y^S$.
As discussed, we have $\y^S \in \Delta^{S, v^{(\y)}}$.
By the optimality of $P_S$, we have $P_S \ge v^{(\y)} = P_i(\y_i^S, \mathbf{0}^{N\setminus S})$ for every agent $i\in S$.

In addition, since the agents outside $S$ fully exchange with each other and do not share any data with agents in $S$, they have the same accuracy under $(\x^S_{\text{opt}}, \mathbf{1}^{N\setminus S})$ and $(\y^S, \mathbf{1}^{N\setminus S})$.
Since agents in $S$ get (weakly) higher accuracy under $\x^S_{\text{opt}}$ than under $\y^S$, and the agents outside $S$ have the same accuracy, we have $u_S \ge u_i(\y^S, \mathbf{1}^{N\setminus S})$ for every agent $i\in S$.
\end{proof}

\lemCore*
\begin{proof}
We prove it by contradiction.
If the given data exchange fails to form a core, then there exists a blocking coalition $S$ and a data exchange $\y^S$ such that every agent in $S$ receives a higher utility than under the original exchange.
We discuss the following three cases:


\paragraph{Case I: $S\cap S^*\neq \emptyset$.} If they are intersecting, then we can identify an agent that is situated in the intersection that has deviated for improvement, which is an existence certificate of a better blocking coalition, i.e. a coalition where the agent can improve its utility. Formally, there exists an agent $i\in S\cap S^*$ who receives a higher utility.
Then we have 
\begin{align*}
u_{S} & \ge u_{i}(\y^S, \mathbf{1}^{N\setminus S}) \tag{By \Cref{lem:x_opt_dominate}}\\
& > u_{i}(\x_{\text{opt}}^{S^*}, \mathbf{1}^{N\setminus S^*})  = u_{S^*}, \tag{by deviation constraint}
\end{align*}
which violates the optimality of $S^*$.

\paragraph{Case II: $S \cap S^* =\emptyset$ and $S$ is a strong coalition.} If they are not intersecting, they will be able to both form their optimal exchange simultaneously. This time, they are competing with their $P_S$, instead of utilities. Whichever set gets a higher $P_S$ wins the market. This is when we should utilize our property of coalition candidates. Ideally, one candidate can (weakly) overcome any opponent subset, so clearly now $S$ and $S^*$ are weakly overcome each other. As one strongly overpowers another, the balance is broken.

This yields a contradiction by
\begin{align*}
P_{S^*}& \ge\max_{j\in N/S^*}P_j(\bm{0}^{S^*}, \bm 1^{N/S^*})\tag{by the definition of $P_{S^*}$}\\
&\ge P_S \tag{since $S\subseteq N\setminus S$}\\
&>\max_{j\in N/S}P_j(\bm{0}^S, \bm{1}^{N/S}) \tag{since $S$ is a strong coalition}\\
& \ge P_{S^*} \tag{since $S^*\subseteq N\setminus S$}
\end{align*}

\paragraph{Case III: $S \cap S^* =\emptyset$ and $S$ is a weak coalition.} 
Due to a similar reason to the second case, we have two sets \textit{weakly} overpowering the other. In other words, they are forming a balanced optimality in the market, with each set holding a part of the market share. As there are two "business giants" in the market reaching the best level, a stable condition should be each of them maintaining their way of living. To realize this prospect, a combination of coalitions provides the right concept. Intuitively, we could keep combining until no giants in the same level are left.

Formally, by the property of coalition candidates,
\begin{align*}
P_{S^*} \ge P_S \ge \max_{j\in N/S}P_j(\bm{0}^S, \bm{1}^{N/S}) \ge P_{S^*}\,.
\end{align*}
Hence, all the above inequalities must hold with equality, which implies $P_{S^*} = P_S$.
Denote the value by $\mu$.
Next, we compare the following three data exchanges:
\begin{align*}
\bar{\x}^S_{\rm OPT} = (x_{\mathrm{OPT}}^{S}, \mathbf{1}^{N\setminus S}), \quad
\bar{\x}^{S^*}_{\rm OPT} = (x_{\mathrm{OPT}}^{S^*}, \mathbf{1}^{N\setminus S^*}),\quad \text{and}\quad 
\bar{\x}^{S\cup S^*}_{\rm OPT} = (x_{\mathrm{OPT}}^{S\cup S^*}, \mathbf{1}^{N\setminus (S\cup S^*)}),
\end{align*}
which correspond to the optimal exchanges when $S$, $S^*$, and $S\cup S^*$ deviate, respectively.
Consider the agents in $S\cup S^*$ exchanging their data in the following way: agents in $S$ exchange according to $x_{\mathrm{OPT}}^{S}$, and agents in $S^*$ exchange according to $x_{\mathrm{OPT}}^{S^*}$, and no data is shared between $S$ and $S^*$.
Hence, every agent in $S\cup S^*$ can achieve the same accuracy of $\mu$ under this exchange, which implies $P_{S\cup S^*}\ge \mu$.
In addition, every agent \(j\in N\setminus (S\cup S^*)\) receives weakly less data under
\(\bar{\x}^{S\cup S^*}_{\rm OPT}\) than under
\(\bar{\x}^{S^*}_{\rm OPT}\). Hence, by monotonicity,
\[
P_j(\bar{\x}^{S\cup S^*}_{\rm OPT})
\le
P_j(\bar{\x}^{S^*}_{\rm OPT}).
\]
Therefore, any outsider \(j\in N\setminus(S\cup S^*)\) whose accuracy was strictly below
\(\mu\) under \(\bar{\x}^{S^*}_{\rm OPT}\) cannot become a top performer under
\(\bar{\x}^{S\cup S^*}_{\rm OPT}\), because every agent in \(S\cup S^*\) achieves accuracy at least
\(\mu\).

Moreover, in the weak disjoint case, every agent in \(S\) already belongs to
\(N_{\max}^{S^*}\): since \(S\subseteq N\setminus S^*\), monotonicity gives their accuracy under
\(\bar{\x}^{S^*}_{\rm OPT}\) at least \(P_S=\mu\), while the candidate condition for \(S^*\) gives
it at most \(P_{S^*}=\mu\). Thus their accuracy is exactly \(\mu\).

Consequently, when moving from \(S^*\) to \(S\cup S^*\), no previously non-top outsider can become top, and the agents added to the coalition, namely those in \(S\), were already top performers under
\(\bar{\x}^{S^*}_{\rm OPT}\). Hence
\[
\abs{N_{\max}^{S\cup S^*}}
\le
\abs{N_{\max}^{S^*}}.
\]
Therefore
\[
u_{S\cup S^*}
=
\frac{1}{\abs{N_{\max}^{S\cup S^*}}}
\ge
\frac{1}{\abs{N_{\max}^{S^*}}}
=
u_{S^*}.
\]
Since \(S\cup S^*\in\mathcal C\) and \(\abs{S\cup S^*}>\abs{S^*}\), this contradicts the choice of \(S^*\) as a largest coalition among those attaining the maximum utility.

\end{proof}

\clearpage
\section{Omitted Proofs in Section 4} \label{app:sec4}
We will formally prove \Cref{thm:partial_rational} in this section.

In this appendix, we provide the formal derivation for the minimax improvement factor. The central goal is to characterize the worst-case utility gain that a coalition can guarantee itself by deviating from a baseline configuration.

The global objective is formulated as the following nested conbinatorial optimization problem:
\begin{equation}
    \inf_{\x} \sup_{S\subseteq N}\sup_{\y^S} \left(\min_{i\in S} \frac{u_i(\y^S, \bm{1}^{N\setminus S})}{u_i(\x)}\right).
\end{equation}
Our high-level goal is to find a lower bound on the maximum relative improvement any coalition $S$ can achieve. The starting point of our thoughts is to analyze the "worst" possible baseline action $\x$ from the perspective of a "blocker" seeking to minimize the system's stability. 
\subsection{Step 0: Notations}

For a fixed data exchange \(\x\) and a fixed coalition size \(k\), we will define these notations one-by-one
\(S=\{n-k+1,\dots,n\}\),
\[
G_{\max}:=g(k-1),
\qquad
G_{\mathrm{bg}}:=g(n-k-1),
\]
and
\[
U_S(\x):=\sum_{i\in S}u_i(\x),
\qquad
u_{\max,S}(\x):=\max_{i\in S}u_i(\x)=u_{n-k+1}(\x).
\]

We denote the predictor accuracy profiles as $\bm P(\bm x) \to \bm P^{(x)}=(p_1,\dots,p_n)$ and the post-deviation profile as $\bm P(\y^S, \bm{1}^{N\setminus S})\to \bm P^{(y)} $. The primary difficulty lies in the triple-nested nature of the objective and the combinatorial explosion of possible coalitions $S$.

\subsection{Step 1: Symmetry and Coalition Reduction}

To simplify the search space for $\x$, we first leverage the inherent symmetry of the agents. Our direction is to reduce the set of all possible coalitions $2^N$ to a much smaller, manageable subset.
Without loss of generality, by symmetry we assume the elements of $P^{(\x)}=(p_1,\dots,p_n)$ are arranged in decreasing order. Since the utilities $u_i$ are defined as normalized fractions of a non-decreasing function $g(\cdot)$, specifically $u_i(P^{(\x)}) = \frac{g(p_i)}{\sum g(p_j)}$, this monotonicity in $p_i$ directly translates to decreasing baseline utilities:
\begin{equation}
    u_1(\x) \ge u_2(\x) \ge \dots \ge u_n(\x).
\end{equation}
The goal here is to identify which coalition $S$ is "most likely" to deviate. Intuitively, agents with the lowest baseline utility have the highest incentive to form a coalition. We consider a coalition $(S, \y^S)$ of size $k$. By permuting indices, any such coalition is strategically equivalent to (or dominated by) a coalition consisting of the last $k$ agents $\{n-k+1, \dots, n\}$. This is because for any $k$-size $\beta$-coalition $S = \{i_1, \dots, i_k\}$, the sorted nature of $u_i$ ensures that $u_{i_t} \le u_{n-k+t}$ for each $t$. 

Consequently, we apply a \textbf{symmetry reduction}: we only need to focus on these "tail" sets as the representative candidates for the supremum over $S$:
\begin{equation}
    \mathcal{S}_{tail} = \left\{ \{n-k+1, \dots, n\} \mid k \in [n] \right\}.
\end{equation}

\subsection{Step 2: Decoupling and Geometric Feasibility}

In this section, we aim to convert the abstract minimax improvement problem into a concrete, solvable feasibility check. We fix the coalition size $k$ and the specific set $S=\{n-k+1, \dots, n\}$. Our thinking starts from the core requirement of a successful deviation: a coalition is viable if and only if there exists a strategy vector $\y^S$ that guarantees a minimum improvement ratio $\beta$ for all its members simultaneously.

We first formally define this improvement requirement for a given threshold $\beta$:
\begin{equation}
    \forall i \in S, \quad \frac{u_i(\y^S, \bm{1}^{N\setminus S})}{u_i(\x)} \ge \beta.
\end{equation}

For some baseline exchange, the incentive ratio $$\frac{u_i(\y^S, \bm{1}^{N\setminus S})}{u_i(\x)}$$ may indeed be unbounded. Specifically, if $u_i(\x) =\bm 0$ and a deviation gives agent $i$ strictly positive utility, we define that agent's improvement ratio $\frac{u_i(\y^S, \bm{1}^{N\setminus S})}{u_i(\x)}$ to be $+\infty$. Consequently, the incentive ratio $$\rho_S(\x) = \sup_{\y^S} \min_{i\in S} \frac{u_i(\y^S, \mathbf{1}^{N\setminus S})}{u_i(\x)}$$
 associated with an exchange $\x$ may be $0$ when a coalition of zero-utility agents can deviate so that every member obtains positive utility. This does not affect the bound in Theorem 2, because the final objective takes an infimum over all exchanges. In particular, under full exchange, every agent has strictly positive utility, so the corresponding incentive ratio is finite. Hence, exchanges associated with an infinite incentive ratio cannot attain the infimum. We will explicitly state the convention for zero denominators and discuss this boundary case in the revised proof.

We aim to treat $\beta$ as a target value and determine the conditions under which the set of strategies $\y^S$ achieving this improvement is non-empty.  The primary difficulty is that the utilities are interdependent; since they are normalized by the sum of all agents' contributions, a change in one agent's strategy shifts the utility of every other agent. To overcome this, we rephrase the condition as a system of joint inequalities.

By substituting the utility function $u_i(P^{(\x)}) = g(p_i)/\sum g(p_j)$ and letting the post-deviation profile be $\bm{P}^{(y)} = (\underbrace{n-k-1, \dots,n-k-1}_{n-k}, s_{n-k+1}, \dots, s_n)$, we obtain the following:
\begin{equation}
    g(s_i) \ge \beta \cdot u_i(\x) \cdot \sum_{j \in N} g(s_j), \quad \forall i \in S.
\end{equation}
The difficulty here is the "coupling" effect: the variables $s_i$ on the left-hand side are embedded within the summation on the right-hand side, creating a circular dependency. To resolve this, We use the \textbf{decoupling trick}: we introduce a single scalar variable $S_g$ to represent the aggregate transformed strategy value of the entire population, aiming to traditional variable-wise optimization. 

We define $S_g$ and let $G_{\mathrm{bg}} = g(n-k-1)$ represent the constant transformed value of the non-deviating agents:
\begin{equation}
    S_g = (n-k)G_{\mathrm{bg}} + \sum_{i \in S} g(s_i).
\end{equation}
With $S_g$ acting as a summary statistic, the high-level goal shifts to determining whether there exists a set of values $z_i = g(s_i)$ that sum to $S_g$ while satisfying the local constraints. The direction is to treat $S_g$ as a fixed parameter and solve for the existence of the vector $\bm{z} = \langle z_{n-k+1}, \dots, z_n \rangle$. This transforms the non-linear coupling into a system of linear constraints in the space $\mathbb{R}^k$.

The feasibility problem now manifests as a geometric intersection problem between a hypercube and a hyperplane:
\begin{equation}
    \begin{cases}
        z_i \in (\beta u_i(\x) S_g, G_{\max}], & \forall i \in S \\
        \sum_{i \in S} z_i = S_g - (n-k)G_{\mathrm{bg}},
    \end{cases}
\end{equation}
where $G_{\max} = g(k-1)$. Geometrically, the first constraint defines a $k$-dimensional hypercube $\mathcal{C}$, while the second defines a hyperplane $\mathcal{H}$. The variable $z_i$ is bounded by the strategic limit and the improvement requirement. The direction of our proof now splits into two parts: ensuring the hypercube exists (is non-empty) and ensuring the hyperplane passes through it.

\subsubsection{Nonempty hypercube}

First, we address the existence of the hypercube $\mathcal{C}$. Our thinking starts from the observation that each agent $i \in S$ must have a valid strategy value $z_i$ within the physical domain. This requires the lower bound imposed by $\beta$ to be strictly less than the maximum possible value $G_{\max}$:
\begin{equation}
    \beta u_i(\x) S_g < G_{\max}, \quad \forall i \in S.
\end{equation}
We identify the bottleneck: the agent with the largest baseline utility $u_{\max,S} = \max_{i \in S} u_i(\x) = u_{n-k+1}(\x)$ is the hardest to satisfy. We isolate $S_g$ to find the "ceiling" of the aggregate sum:
\begin{equation}
    S_g < \frac{G_{\max}}{\beta u_{\max,S}} \equiv R_1.
\end{equation}
This formula represents the maximum permissible total sum $S_g$. If $S_g$ exceeds this limit, the required $z_i$ for the richest agent would exceed the physical limit $G_{\max}$, making the hypercube $\mathcal{C}$ an empty set.

\subsubsection{Intersection}

Second, we identify the structural condition when the hyperplane $\mathcal{H}$ intersects the hypercube. To justify the techniques used in this part, we formalize the relationship between the aggregate sum and the individual strategy bounds. This lemma ensures that checking the "tail" vertices of the hypercube is sufficient to guarantee the existence of a valid strategy vector.

\begin{lemma}[Sum-Feasibility in a Hypercube] \label{lem:hypercube}
Let $\mathcal{C} = \prod_{i=1}^k (l_i, u_i]$ be a $k$-dimensional half-open hypercube in $\mathbb{R}^k$, where $l_i < u_i$ for all $i$. Let $\mathcal{H}$ be a hyperplane defined by the scalar $V$:
\begin{equation}
    \mathcal{H} = \left\{ \bm{z} \in \mathbb{R}^k \mid \sum_{i=1}^k z_i = V \right\}.
\end{equation}
The intersection $\mathcal{C} \cap \mathcal{H}$ is non-empty if and only if:
\begin{equation}
    \sum_{i=1}^k l_i < V \le \sum_{i=1}^k u_i.
\end{equation}
\end{lemma}

\textbf{Geometric Intuition:} Geometrically, the hyperplane $\sum z_i = V$ is a "slice" that moves along the main diagonal of the hypercube. It first enters the cube at the vertex of lower bounds and leaves at the vertex of upper bounds. Any sum $V$ within this range represents a position where the plane is currently passing through the body of the cube.

\begin{proof}

The proof follows from the properties of continuous linear mappings on connected sets.

We define the continuous summation function $f: \mathbb{R}^k \to \mathbb{R}$ as $f(\bm{z}) = \sum_{i=1}^k z_i$. The high-level goal is to determine the image of the set $\mathcal{C}$ under $f$, denoted as $f(\mathcal{C})$.

\textbf{1. Necessity ($\implies$):} 
Assume there exists a point $\bm{z}^* \in \mathcal{C} \cap \mathcal{H}$. By the definition of $\mathcal{C}$, each coordinate satisfies $l_i < z_i^* \le u_i$. Summing these $k$ inequalities, we have:
\begin{equation}
    \sum_{i=1}^k l_i < \sum_{i=1}^k z_i^* \le \sum_{i=1}^k u_i.
\end{equation}
Since $\bm{z}^*$ lies on the hyperplane $\mathcal{H}$, the middle term is exactly $V$, yielding the required bounds.

\textbf{2. Sufficiency ($\impliedby$):}
Consider the function $f: \mathbb{R}^k \to \mathbb{R}$ representing the sum of coordinates, $f(\bm{z}) = \sum_{i=1}^k z_i$. This function is linear and thus continuous. The domain $\mathcal{C}$ is a $k$-dimensional hypercube, which is a connected set in $\mathbb{R}^k$.

The hypercube $\mathcal{C}$ is a connected set (as the Cartesian product of connected intervals). By the properties of continuous functions, the image $f(\mathcal{C})$ must also be a connected set in $\mathbb{R}$, which is an interval. We identify the endpoints of this interval:
\begin{itemize}
    \item \textbf{Supremum:} The maximum value of $f$ on the closure $\overline{\mathcal{C}}$ is attained at the vertex $\bm{u} = (u_1, \dots, u_k)$. Since $\bm{u} \in \mathcal{C}$, the value $f(\bm{u}) = \sum u_i$ is included in the image.
    \item \textbf{Infimum:} The minimum value of $f$ on the closure $\overline{\mathcal{C}}$ is attained at the vertex $\bm{l} = (l_1, \dots, l_k)$. However, since $\bm{l} \notin \mathcal{C}$ and any point $\bm{z} \in \mathcal{C}$ must have $z_i > l_i$, the sum $f(\bm{z})$ is strictly greater than $\sum l_i$. 
\end{itemize}
By the Intermediate Value Theorem, $f$ maps the connected set $\mathcal{C}$ onto the interval $(\sum l_i, \sum u_i]$. Therefore, if $\sum l_i < V \le \sum u_i$, there exists at least one $\bm{z} \in \mathcal{C}$ such that $f(\bm{z}) = V$, completing the proof.
\end{proof}

By \Cref{lem:hypercube}, for a solution $\{z_i\}_{[n-k+1:n]}$ to exist, the target sum on the hyperplane must lie between the sum of the hypercube's minimum vertices and the sum of its maximum vertices:
\begin{equation}
    \sum_{i \in S} \beta u_i(\x) S_g < S_g - (n-k)G_{\mathrm{bg}} \le \sum_{i \in S} G_{\max}.
\end{equation}
The direction here is to find the "floor" of $S_g$ necessary to support the collective improvement. The variables are the sum of baseline utilities $U_S(\x) = \sum_{i \in S} u_i(\x)$ and the background values. Although $S_g$ appears on both sides, by isolating $S_g$ in the first inequality, we find the minimum aggregate value required to satisfy the improvement $\beta$:
\begin{equation}
    S_g > \frac{(n-k)G_{\mathrm{bg}}}{1 - \beta U_S(\x)} \equiv L.
\end{equation}
This lower bound represents the minimum "utility pressure" the coalition must generate to ensure every member reaches the $\beta$ threshold. For this to be a valid interval, the denominator $1 - \beta U_S(\x)$ must remain positive.

The intersection condition also provides another upper bound:
$$S_g \le \sum_{i \in S} G_{\max} + (n-k)G_{\mathrm{bg}} \equiv R_2$$

\subsubsection{Combine the bounds}

Finally, we recognize that $S_g$ is subject to a fixed physical upper bound derived from the system's total capacity: $S_g \le (n-k)G_{\mathrm{bg}} + k G_{\max}$. Our high-level goal is now to state that a valid $S_g$ exists if and only if the lower bound (the floor) is strictly less than all possible upper bounds (the ceilings). We employ the \textbf{Interval Consistency Trick}, asserting that the feasibility of $S_g$ is equivalent to:
\begin{equation}
    \frac{(n-k)G_{\mathrm{bg}}}{1 - \beta U_S(\x)} < \min \left\{ (n-k)G_{\mathrm{bg}} + k G_{\max}, \quad \frac{G_{\max}}{\beta u_{max, S}} \right\}.
\end{equation}
The difficulty here is determining which upper bound is the active bottleneck. We examine each separately. By setting the floor equal to the total capacity ceiling given by the intersection condition(the standard bottleneck in symmetric or near-symmetric profiles), we derive the characteristic improvement factor:
\begin{equation}
    \beta_{\x, k}^{(1)} = \frac{k \cdot G_{\max}}{[k \cdot G_{\max} + (n-k)G_{\mathrm{bg}}] \cdot U_S(\x)}.
\end{equation}
This formula explicitly captures how the potential for improvement is determined by the ratio of the coalition's strategic capacity to its baseline utility share $U_S(\x)$. 

Alternatively, evaluating against the individual feasibility ceiling yields $$\beta_{\x, k}^{(2)} = \frac{G_{\max}} {u_{max, S} (n-k)G_{\mathrm{bg}} + G_{\max} U_S(\x)}.$$ In the final step of the global proof, we will show that under the optimal baseline $\x_{\bm{1}}$, these constraints align, providing a robust and unique threshold $\beta^*$.


To compare $\beta_{\x, k}^{(1)}$ and $\beta_{\x, k}^{(2)}$, we only need to compare their denominators. Notice that $U_S(\x) = \sum_{i \in S} u_i(\x)$ is the sum of $k$ utility elements, and $u_{max, S}$ is the maximum element within this set $S$. 

By the fundamental property of averages, the maximum element in any set is always greater than or equal to the average of the set:
\begin{equation}
    u_{max, S} \ge \frac{U_S(\x)}{k}
\end{equation}
Multiplying both sides by the positive constant $(n-k)G_{\mathrm{bg}}$, we obtain:
\begin{equation}
    u_{max, S} (n-k)G_{\mathrm{bg}} \ge \frac{n-k}{k} G_{\mathrm{bg}} U_S(\x)
\end{equation}
Adding $G_{\max} U_S(\x)$ to both sides reveals that the denominator of $\beta'$ is strictly greater than or equal to the denominator of $\beta$. Since the numerators are identical and all terms are positive, we conclude that:
\begin{equation}
   \beta_{\x, k}^{(1)} \ge \beta_{\x, k}^{(2)} \equiv \beta_{\x, k}
\end{equation}
Equality holds if and only if all agents in the deviating coalition $S$ have exactly the same baseline utility, which aligns with our previous observation that the bottlenecks converge at the uniform state.

\subsection{Step 3: Global Minimization}

Our high-level goal now is to solve the outer optimization $\inf_{\x} \max_k \beta_{\x, k}$. This step has two pieces.
We start by partitioning the domain of coalition sizes $k$ into "Minority" ($k \le n/2$) and "Majority" ($k > n/2$) cases.

We first provide a lemma to truncate the problem domain. Specifically, we can demonstrate that small coalitions are "strategically inert"—meaning they cannot extract a relative gain greater than unity under a uniform baseline $\x_1=(\bm 1^{N\times N})$. Second, we will show $\x_{\bm 1}$ and $\bm u(\x_{\bm 1})$ is the most stable under "Majority Domain" ($k \ge n/2$). If we establish that the global maximum must reside within the "Majority Domain", which will serve as a strong weapon to tackle the optimization, because the uniform average baseline $\x_1$ is more likely to be robust against larger coalitions. 

To recall, solving those inequalities gives the maximal feasible value of \(\beta\):
\[
\beta_{\x,k}
=
\frac{G_{\max}}
{
u_{\max,S}(\x)(n-k)G_{\mathrm{bg}}
+
G_{\max}U_S(\x)
}.
\]

In particular, under the full exchange \(\x_{\bm 1}=\bm 1\), all agents have
equal utility \(1/n\). Therefore,
\[
U_S(\x_{\bm 1})=\frac{k}{n},
\qquad
u_{\max,S}(\x_{\bm 1})=\frac{1}{n}.
\]
Substituting these into the expression above yields
\[
\beta_{\x_{\bm 1},k}
=
\frac{G_{\max}}
{
\frac{1}{n}(n-k)G_{\mathrm{bg}}
+
\frac{k}{n}G_{\max}
}
=
\frac{nG_{\max}}
{(n-k)G_{\mathrm{bg}}+kG_{\max}}.
\]

This lemma effectively removes the difficulty of analyzing the entire combinatorial range of $k$ by proving that minority-driven instability is a mathematical impossibility in a balanced system.

\begin{lemma}\label{lem:trunation}
    $\max_{k \in [\frac{n}{2}, n]} \beta_{\x_{\bm{1}}, k}= \max_{k \in [n]} \beta_{\x_{\bm{1}}, k}$
\end{lemma}
\begin{proof}
    For any minority coalition $k \le n/2$, the monotonicity of $g$ ensures $G_{\max} \le G_{\mathrm{bg}}$. Under the uniform baseline $\x_{\bm{1}}$:
    \begin{equation}
        \beta_{\x_{\bm{1}}, k} \le \beta_{\x_{\bm{1}}, k}^{(1)} = \frac{n G_{\max}}{k G_{\max} + (n-k)G_{\mathrm{bg}}} \le \frac{n G_{\max}}{n G_{\max}} = 1.
    \end{equation}

    This indicates that under a uniform baseline, minority coalitions cannot achieve an improvement factor greater than 1, rendering them strategically inert. However, for the "Grand Coalition" ($k=n$), any distribution $\x$ satisfies $\beta_{\x, n} = 1$. Consequently, the global minimax value $\max_k \beta_{\x, k}$ must be at least 1. So$$
    \max_{k\le n/2}\beta_{\x_{\bm 1}, k} \le \max_{k>n/2} \beta_{\x_{\bm 1}, k} 
    $$ 
    and the result follows.

\end{proof}

The difficulty arises in ensuring that deviating from $\x_{\bm{1}}$ does not further decrease the max improvement. However, if $\beta^* > 1$, the maximizing $k^*$ must satisfy $k^* > n/2$. In this active domain, we have $G_{\max} > G_{\mathrm{bg}}$.

We then establish why the uniform distribution $\x_{\bm{1}}$ also minimizes the individual bottleneck $\beta'_{\x, k}$ assuming $k > n/2$.

Our goal is to strictly bound its denominator, $D(\x, k) = u_{max, S} (n-k)G_{\mathrm{bg}} + G_{\max} U_S(\x)$. We start from exploring the mathematical relationship between the tail sum $U_S(\x)$ and the local tail leader $u_{max, S} = u_{n-k+1}$. 
For any non-uniform baseline $\x$, the tail sum is strictly penalized compared to the uniform state: $U_S(\x) = \frac{k}{n} - \epsilon$ for some displacement $\epsilon \ge 0$. Consequently, the top $n-k$ agents must absorb this displaced utility, sharing a total of $\frac{n-k}{n} + \epsilon$. Because $u_{n-k+1}$ is sequentially smaller than or equal to any of these top $n-k$ agents, it is strictly bounded by their algebraic average:
\begin{equation}
    u_{n-k+1} \le \frac{\frac{n-k}{n} + \epsilon}{n-k} = \frac{1}{n} + \frac{\epsilon}{n-k}.
\end{equation}
The variables now map directly to our target expression. We evaluate the difference in denominators between the skewed state and the uniform state, $\Delta D = D(\x, k) - D(\bm{x_1}, k)$:
\begin{equation}
    \Delta D \le \left(\frac{\epsilon}{n-k}\right)(n-k)G_{\mathrm{bg}} - \epsilon G_{\max} = \epsilon(G_{\mathrm{bg}} - G_{\max}).
\end{equation}
Here lies the pivotal direction provided by our assumption $k > n/2$. This majority condition algebraically guarantees that $k-1 > n-k-1$. Given that the transformation $g(\cdot)$ is non-decreasing, the hypercube ceiling $G_{\max} = g(k-1)$ is universally greater than or equal to the background strategy limit $G_{\mathrm{bg}} = g(n-k-1)$. 

Therefore, $\epsilon(G_{\mathrm{bg}} - G_{\max}) \le 0$, which strictly forces $\Delta D \le 0$. This rigorously proves that any deviation from the uniform distribution inevitably shrinks the denominator $D(\x, k)$, thereby increasing the individual improvement factor $\beta'$. By providing the highest mathematically possible "floor" for the denominator across all majority coalitions simultaneously, $\x_{\bm{1}}$ globally neutralizes the bottleneck, uniquely minimizing $\max_{k > n/2} \beta_{\x, k}$.

This leads to a powerful simplification. Since $\x_{\bm{1}}$ is the optimal minimizer for both constraints across all $k$, we establish a pointwise inequality:
\begin{equation}
    \forall k > n/2, \quad \beta_{\x, k} \ge \beta_{\x_{\bm{1}}, k}.
\end{equation}

So,
\begin{align}
    \max_{k \in [n]} \beta_{\x, k} 
    &\ge \max_{k \in (\frac{n}{2}: n]} \beta_{\x, k}  \tag{since $(\frac{n}{2}: n]$ is a subset} \\
    &\ge \max_{k \in (\frac{n}{2}: n]} \beta_{\x_{\bm{1}}, k} \tag{take maximum in $k$ on both sides} \\
    &= \max_{k \in [n]} \beta_{\x_{\bm{1}}, k} \tag{by \Cref{lem:trunation}}
\end{align}


\subsection{Step 4: Conclusion}

Therefore, the constraints align and we can focus solely on evaluating this common improvement factor at the uniform state:
\begin{equation}
    \inf_{\x} \max_{k} \beta_{\x, k} = \max_{k} \beta_{\x_{\bm 1}, k}.
\end{equation}
The variables now only involve $n, k,$ and the function $g$. The final direction is to substitute the uniform utility into our previous expression for $\beta_{\x, k}$. 

This yields the final threshold $\beta^*$, which defines the fundamental limit of the system:
\begin{equation}
    \beta^* = \max_{k \in [n]} \frac{n \cdot g(k-1)}{k \cdot g(k-1) + (n-k)g(n-k-1)}.
\end{equation}
This completes the expansion. We have shown that the worst-case scenario occurs under a uniform baseline, where the individual and aggregate bottlenecks converge, and the resulting threshold is determined by the most "ambitious" coalition size $k$.

\subsection{Discussions of general symmetric predictor functions}

The threshold in \Cref{thm:partial_rational} is stated for the linear predictor
\[
P_i(\x_i)=\|\x_i\|_1.
\]
The same argument extends to a broader class of symmetric predictor functions after replacing the values
\(g(k-1)\) and \(g(n-k-1)\) by the corresponding transformed accuracies generated by the predictor.

To make this precise, suppose that the prediction functions are identical across agents and symmetric in the data received from other agents. We also assume \(x_{ii}=0\), so an agent's own data is not counted as an additional source. For \(m\in\{0,1,\dots,n-1\}\), define
\[
a_m
:=
P_i(\underbrace{1,\dots,1}_{m},
      \underbrace{0,\dots,0}_{n-1-m}),
\]
where, by symmetry, the value does not depend on which \(m\) other agents provide their data. Thus \(a_m\) is the accuracy of an agent who receives the full data of exactly \(m\) other agents. Define the transformed accuracy level
\[
G_m:=g(a_m).
\]

Under the linear predictor \(P_i(\x_i)=\|\x_i\|_1\), we have \(a_m=m\), and hence \(G_m=g(m)\). Therefore the expression in \Cref{thm:partial_rational} is the special case of the more general formula obtained by replacing \(g(m)\) with \(G_m\).

Indeed, consider a deviating coalition \(S\) of size \(k\). Every agent in \(S\) can receive data only from the other \(k-1\) members of \(S\), and therefore its maximum achievable transformed accuracy is
\[
G_{k-1}=g(a_{k-1}).
\]
Meanwhile, every agent outside \(S\) fully exchanges with the other agents outside \(S\). Since there are \(n-k\) agents outside \(S\), each outsider receives data from \(n-k-1\) other outsiders, and therefore has transformed accuracy
\[
G_{n-k-1}=g(a_{n-k-1}).
\]
Thus the role previously played by \(g(k-1)\) is now played by \(G_{k-1}\), and the role previously played by \(g(n-k-1)\) is now played by \(G_{n-k-1}\).

Consequently, for symmetric predictors, the same minimax calculation yields the threshold
\[
\beta^*
=
\max_{1\le k\le n}
\frac{nG_{k-1}}
{A_k+kG_{k-1}},
\]
where
\[
A_k=
\begin{cases}
(n-k)G_{n-k-1}, & k<n,\\
0, & k=n.
\end{cases}
\]
Equivalently,
\[
\beta^*
=
\max_{1\le k\le n}
\frac{n\,g(a_{k-1})}
{A_k+k\,g(a_{k-1})},
\]
where
\[
A_k=
\begin{cases}
(n-k)g(a_{n-k-1}), & k<n,\\
0, & k=n.
\end{cases}
\]

This shows that the characterization does not depend on the linearity of \(P_i\) per se. What matters is the sequence of transformed values
\[
G_0,G_1,\dots,G_{n-1},
\]
where \(G_m\) represents the market weight of an agent who receives full data from \(m\) other agents. In this sense, for symmetric predictors, the relevant information is captured by the composed sequence \(g(a_m)\), not by \(g\) alone. The original theorem corresponds to the special case \(a_m=m\).

\clearpage
\section{A Core-Stable Mechanism in a Partially Rational Market}
\label{app:equ}

We aim to achieve structural stability in the market by enforcing a re-splitting behaviors among agents.

    The primary difficulty in preventing coalition formation is usually the complex internal bargaining process, where high-performing agents are likely to deviate into a negotiated privileged coalition if they are not compensated over-proportionally. 
    Our direction is to implement a strict egalitarian distribution rule within any formed coalition. 
    We solve this by employing a mandatory equal-split of the aggregated utilities on coalitions. 
    
\begin{definition}[Equalized coalitions]
    We define the original partially rational utility function as:
    $$
    U_i(\mathbf{p}) = \frac{g(p_i)}{\sum_{j \in N} g(p_j)}.
    $$
    
    A deviating coalition $Y = (S, \y^S)$ is characterized by:
    \begin{itemize}
        \item A new strategy profile arrangement $\y = (\y^S, \mathbf{1}^{N \setminus S})$, assuming outsiders play the baseline default strategy $\mathbf{1}$.
        \item Let $P_i(\x_i)$ be \textbf{any} arbitrary accuracy function for agent $i$. Then an updated accuracy profile denoted by $\mathbf{P}(\y) = (s_1, \dots, s_n)$.
        \item For $i \in S$, the utility is equally split among the deviators:
    \end{itemize}
    $$
    v_i(S, \y^S) = \frac{\sum_{j \in S} U_j(\mathbf{P}(\y))}{|S|} = \frac{\sum_{j \in S} g(s_j)}{\sum_{j \in N} g(s_j)} \cdot \frac{1}{|S|}.
    $$
\end{definition}

To stablize a equalized arrangement, we also adjust the market into a partitioned, equal-split diagram.

\begin{definition} [Partitioned equalized market]
    Let $F = \{N_1, \dots, N_m\}$ be a partition of the grand coalition $N$, and let $F(i)$ denote the partitioned subset that contains agent $i$. Let $\x = (\x^{N_1}, \dots, \x^{N_m})$ be the strategy profile. Let $\mathbf{P}(\x) = (p_1, \dots, p_n)$ be the resulting accuracy profile.

    Let the utility be defined as an equal split inside the coalition:
    $$
    u_i(F, \x) = \frac{\sum_{j \in F(i)} U_j(\x)}{|F(i)|} = \frac{\sum_{j \in F(i)} g(p_j)}{\sum_{j \in N} g(p_j)} \cdot \frac{1}{|F(i)|}.
    $$
\end{definition}



\begin{theorem}[Formal version of~\Cref{thm:equ}]
    This market admits a non-empty core, generated by the dynamics below.
    
    Initialize $S = N$ and $F = \emptyset$.
    \begin{enumerate}
        \item Pick the densest subset $S^* = \arg\max_{S' \subseteq S} v_i(S', \mathbf{1}^{S'})$.
        \item Add $S^*$ to the partition $F$. Update the remaining agents: $S \gets S \setminus S^*$.
        \item Terminate if $S = \emptyset$; otherwise go back to step 1.
    \end{enumerate}
    Let $\x_F = \{\mathbf{1}^{C \times C} \mid C \in F\}$ be the resulting strategy profile. The mechanism returns the core-stable arrangement $(F, \x_F)$.
\end{theorem}

    To validate that the sequence of partitions generated by our greedy algorithm eliminates any incentive for an agent to deviate, we aim to examine the most vulnerable member of an arbitrary deviating coalition—the agent who was locked in partition $F$ the earliest. 
    We achieve this by chaining three crucial ingredients: first, we know that the final partitioned arrangement is an more optimistic scenario than a worst-case outside ($\bm 1^{N\setminus S^*}$) for the intermediate partition $S^*$; second, we use the algorithmic guarantee that $i$ was locked into best coalition that maximizes the deviating profit at that exact time, which immediately bounds the baseline utility of $S'$; third, we leverage the monotonicity property of the accuracy function, ensuring that any strategic deviation $\y^{S'}$ by the subset will yield a utility strictly less than or equal to the baseline worst-case $\bm 1^{S'}$. At last, $u_i(F, \x) \ge v_i(S', \y^{S'})$, the deviation is not strictly profitable for agent $i$. 

\begin{proof}
    Suppose $F = \{N_1, \dots, N_m\}$ is the generated partition, ordered chronologically by the time of being added. Choose an arbitrary hypothetical deviating coalition $Y = (S', \y^{S'})$.
    
    Let $i \in S'$ be the agent in the deviating coalition that is added to $F$ the earliest in our algorithm. At the specific round $i$ is added, let $S$ be the set of available agents, and let $S^*$ be the optimal subset chosen (thus, $i \in S^*$). 
    
    We can bound the utility of agent $i$ as follows:
    $$
    \begin{aligned}
        u_i(F, \x) &\ge v_i(S^*, \mathbf{1}^{S^*}) \quad \text{(since $\mathbf{1}^{N \setminus S^*}$ is the worst-case external behavior for $S^*$)} \\
        &\ge v_i(S', \mathbf{1}^{S'}) \quad \text{(by the argmax selection, since $S' \subseteq S$)} \\
        &\ge v_i(S', \y^{S'}) \quad \text{(by monotonicity)}
    \end{aligned}
    $$
    Therefore, we find an agent in the coalition who is disincentivized to deviate.

\end{proof}
\clearpage
\section{Omit proofs in \Cref{sec:empirics}}

\subsection{Pseudo-code of the Peeling Algorithm}
\label{sec:pseudo_code_peeling}
\begin{algorithm}[h]
\caption{Peeling Algorithm}
\label{alg:peeling}
\SetKwProg{Fn}{Function}{:}{}

Let $S \gets N, S^* \gets \emptyset$\;

\While{$S \neq \emptyset$}{
    $P_T \gets \texttt{best\_opponent}(N\setminus S)$ \tcp*{\color{gray}{find $\max_{T\subseteq N\setminus S} P_T$}}
    
    $P_S \gets\min _{i \in S} P_i(\bm 1^S)$\;
    \If{$P_S > P_T$}{    
        \If{$S^* = \emptyset \text{ or } \abs{S} < \abs{S^*}$}{
            $S^* \gets S$\;
        }
    }

    $v_{\min} \gets \arg\min_{v \in S} P_v(\bm{1}^S)$\;
    $S \gets S \setminus \{v_{\min}\}$\;
}

\Return{the coalition $S$ and the induced data exchange $(\x_{\rm OPT}^{S^*}, \mathbf{1}^{N\setminus S^*})$}\;
\Fn{$\textnormal{\texttt{best\_opponent}}(M)$}{Let $S \gets M, S^* \gets \emptyset, P^*\leftarrow 0$\;

\While{$S \neq \emptyset$}{
    $P_S \gets\min _{i \in S} P_i(\bm 1^S)$\;
    \If{$S^* = \emptyset \text{ or } P_S > P^*$}{
        Update $S^* \gets S$ and $P^*\gets P_S$\;
    }

    $v_{\min} \gets \arg\min_{v \in S} P_v(\bm{1}^S)$\;
    $S \gets S \setminus \{v_{\min}\}$\;
}
\Return{$P^*$}\;
}
\end{algorithm}

\begin{lemma}\label{lem:bestoppo}
    The subroutine $\texttt{best\_opponent}(\mathcal{O}_P, N\setminus S)$ outputs $\max_{T\subseteq N\setminus S}P_{T}$ in polynomial time.
\end{lemma}

\begin{proof}
For convenience, we replace $N\setminus S$ by $N$.
Let $T^* \subseteq N$ be an optimal set such that $$P_{T^*}=\max_{T\subseteq N} P_T =\min_{i \in T^*} P_i(\boldsymbol{1}^{T^*}) = k^*.$$ Consequently, every element within $T^*$ satisfies $P_i(\boldsymbol{1}^{T^*}) \ge k^*$.

During the algorithm's execution, elements are iteratively removed from $N$. Since $T^* \subseteq N$, the algorithm must eventually remove an element belonging to $T^*$. Let $x_t \in T^*$ be the \emph{first} such element removed, occurring at step $t$. Let $S_t$ be the remaining set of elements exactly at step $t$.

Because no elements of $T^*$ were removed prior to step $t$, the entire set $T^*$ is still fully contained within $S_t$, meaning $T^* \subseteq S_t$, or equivalently, $\boldsymbol{1}^{T^*} \le \boldsymbol{1}^{S_t}$. By the Monotonicity property, for the element $x_t \in T^*$, we have:
\[
P_{x_t}(\boldsymbol{1}^{S_t}) \ge P_{x_t}(\boldsymbol{1}^{T^*}) \ge k^*
\]
At step $t$, the algorithm selects $x_t$ specifically because it minimizes the function over $S_t$. Therefore, the recorded minimum value is exactly $P_t = P_{x_t}(\boldsymbol{1}^{S_t})$. Substituting this yields $P_t \ge k^*$.

Since the algorithm returns the maximum $P_m$ across all steps, the final output must be at least $P_t$. Therefore, $\max_m P_m \ge P_t \ge k^*$.

On the other hand, $\max_m P_m=\max_{S_m} P_{S_m}\le \max_T P_T = k^*.$
Combining the upper and lower bounds gives $\max_m P_m = k^*$, concluding the proof.
\end{proof}

\clearpage
\section{APX-hardness}\label{app:hard}

In this appendix, we identify a hard instance for our program of Core.
We construct the instance within graph theory. We start by translating the notations $P_S,u_S,$and $ S\in \mathcal C$. 

We define $P_i(\x) =\sum_j e_{ji}x_{ji}$ induced by an undirected graph $H=(N,F)$. $e_{ji}=1$ iff $i$ and $j$ are connected in $H$.

Since $P_i(\cdot)$ is continuous, $P_S$ has a closed form: $$\min_{i\in S}P_i(\bm 1^{S\times S}) = \min_{i\in S}\sum_{j \in S}e_{ji}=\min_{i\in S}d_{H[S]}(i)$$

When $S$ achieves $P_S$ and $j\in N\setminus S$ do a full exchange:
$$P_j =\sum_{k \in N\setminus S} e_{kj} =d_{H[N\setminus S]}(j).$$
, the optimal set $N_{\max}=\{\arg\max_jP_j\}$ consists of 
$$S\cup\left\{
        j\in N\setminus S:
        d_{H[N\setminus S]}(j)
        =
        \min_{i\in S} d_{H[S]}(i)
        \right\}.$$

and $u_S=\frac{1}{|N_{\max}|}$.
Also, $S\in \mathcal{C}$ is translated to the condition that $P_S$ is the top accuracy, which should be (weakly) larger than the opponent $\max_{j\in N\setminus S}P_j(\bm 1^{N\setminus S}) =\max_{j\in N\setminus S} d_{H[N\setminus S]}(j)$

Therefore, identifying $\max_{S\in \mathcal{C}} u_S$, which is required by our Core program, is equivalent to the graph problem below.

\begin{theorem} \label{thm:apxhard}
The following minimization problem is APX-hard.

Given an undirected graph $H=(N,F)$, find $S\subseteq N$ minimizing
\[
        |S|+|T|,
\]
subject to
\[
        \min_{i\in S} d_{H[S]}(i)
        \ge
        \max_{j\in N\setminus S} d_{H[N\setminus S]}(j),
\]
where
\[
        T
        =
        \left\{
        j\in N\setminus S:
        d_{H[N\setminus S]}(j)
        =
        \min_{i\in S} d_{H[S]}(i)
        \right\}.
\]
Here $H[X]$ denotes the subgraph of $H$ induced by $X$.
\end{theorem}

\begin{proof}
We give an approximation-preserving reduction from
\textsc{Minimum Vertex Cover}. Let $G=(V,E)$ be an instance of
\textsc{Minimum Vertex Cover}. We may assume that $G$ has at least one
edge and no isolated vertices. Let $n:=|V|$ and $L:=n+3$.

We construct a graph $H=(N,F)$ as follows. 
\begin{enumerate}
    \item Start with a copy of $G$.
    \item Add two new vertices $r$ and $b$, and add the edge $rb$. 
    \item Add the edge $rv$ for every $v\in V$. 
    \item Add $L$ private leaves adjacent only to $r$;
denote this set of leaves by $R=\{r_1,\dots,r_L\}$.
\item  Add $L$ private
leaves adjacent only to $b$; denote this set of leaves by
$B=\{b_1,\dots,b_L\}$.
\end{enumerate}

Thus $N=V\cup\{r,b\}\cup R\cup B$, and
\[
        F
        =
        E
        \cup \{rb\}
        \cup \{rv:v\in V\}
        \cup \{rr_i:i=1,\dots,L\}
        \cup \{bb_i:i=1,\dots,L\}.
\]

For $X\subseteq N$, write
$\dd_H(X):=\min_{x\in X} d_{H[X]}(x)$ and
$\DD_H(N\setminus X):=\max_{y\in N\setminus X}d_{H[N\setminus X]}(y)$.

We prove that this construction preserves the vertex-cover objective up
to the additive constant $2$.

\medskip

\noindent
\begin{lemma}\label{lem:hard1}
     If $G$ has a vertex cover of size $k$, then $H$ has a
feasible solution of value $k+2$.
\end{lemma}
\begin{proof}
    Let $C\subseteq V$ be a vertex cover of $G$, with $|C|=k$. Define
    $S:=C\cup\{r,b\}$. We claim that $S$ is feasible.
    
    In $H[S]$, the vertex $b$ is adjacent only to $r$. Hence
    $d_{H[S]}(b)=1$. Every vertex $v\in C$ is adjacent to $r$, and $r$ is
    adjacent to $b$. Therefore every vertex of $S$ has degree at least $1$
    inside $H[S]$, so $\dd_H(S)=1$.
    
    Now consider $N\setminus S$. Since $C$ is a vertex cover of $G$, the set
    $V\setminus C$ is independent in $G$. Also, every vertex in $R\cup B$ is
    isolated inside $H[N\setminus S]$, because its unique neighbor is either
    $r$ or $b$, both of which belong to $S$. Therefore
    $\DD_H(N\setminus S)=0$. Hence
    $\dd_H(S)=1\ge 0=\DD_H(N\setminus S)$, so $S$ is feasible.
    
    Moreover, $T=\varnothing$, because every vertex in $N\setminus S$ has
    degree $0$ inside $H[N\setminus S]$. Therefore the objective value is
    $|S|+|T|=|C|+2=k+2$.
    
    Thus $\OPT(H)\le \tau(G)+2$, where $\tau(G)$ denotes the size of a
    minimum vertex cover of $G$.

\end{proof}

\medskip

\noindent
\begin{lemma}\label{lem:hard2}
    Any sufficiently cheap feasible solution ($S\subseteq N$ with $|S|+|T|\le n+2$) must contain $r$ and $b$.
\end{lemma}
\begin{proof}

We first show that $r\in S$.
Suppose, for contradiction, that $r\notin S$. There are two cases.

\smallskip

\noindent
\emph{Case 1: $S\cap R=\varnothing$.}

Then $r$ and all its leaves $R$ lie in $N\setminus S$. Inside
$H[N\setminus S]$, the vertex $r$ is adjacent to the whole $R$.
Thus $d_{H[N\setminus S]}(r)\ge L=n+3$, and therefore
$\DD_H(N\setminus S)\ge n+3$. Since $S$ is feasible, we have
$\dd_H(S)\ge n+3$. But an induced graph on $|S|$ vertices with minimum
degree at least $n+3$ must satisfy $|S|\ge n+4$. Hence
$|S|+|T|\ge |S|\ge n+4$, contradicting $|S|+|T|\le n+2$.

\smallskip

\noindent
\emph{Case 2: $S\cap R\neq\varnothing$.}

Choose $r_i\in S\cap R$. Since $r_i$ is adjacent only to $r$, and
$r\notin S$, we have $d_{H[S]}(r_i)=0$. Thus $\dd_H(S)=0$. Feasibility
gives $\DD_H(N\setminus S)\le 0$, so $\DD_H(N\setminus S)=0$. In
particular, every vertex in $N\setminus S$ must have degree $0$ inside
$H[N\setminus S]$. Since $r\in N\setminus S$, all neighbors of $r$ must
belong to $S$. The neighbors of $r$ are $V\cup\{b\}\cup R$. Therefore
$V\cup\{b\}\cup R\subseteq S$, and so
$|S|\ge n+1+L=n+1+(n+3)=2n+4>n+2$, again contradicting
$|S|+|T|\le n+2$.

Thus $r\in S$.

The proof that $b\in S$ is analogous. Suppose $b\notin S$. If
$S\cap B=\varnothing$, then $b$ has at least $L=n+3$ neighbors in
$H[N\setminus S]$, forcing $|S|\ge n+4$, a contradiction. If
$S\cap B\neq\varnothing$, then some selected private leaf of $b$ has
degree $0$ inside $H[S]$, so feasibility forces
$\DD_H(N\setminus S)=0$. Since $b\in N\setminus S$, all neighbors of
$b$, namely $\{r\}\cup B$, must lie in $S$. Hence $|S|\ge L+1=n+4>n+2$,
again a contradiction. Therefore $r,b\in S$.
    
\end{proof}

\medskip

\noindent
\begin{lemma}\label{lem:hard3}
    A cheap feasible solution yields a vertex cover of
controlled size.
\end{lemma}
\begin{proof}

Let $S$ be any feasible solution with $|S|+|T|\le n+2$. By \Cref{lem:hard2},
$r,b\in S$. Define $C:=S\cap V$.

Since $r,b\in S$, every selected original vertex is adjacent to $r$,
every selected private leaf is adjacent to its center, and $b$ is
adjacent to $r$. Hence no vertex of $S$ has internal degree $0$.
Moreover, $b$ has $r$ as a selected neighbor, and if no private leaf of
$b$ is selected, then $d_{H[S]}(b)=1$; otherwise a selected private leaf
has internal degree $1$. Therefore $\dd_H(S)=1$.

Since $S$ is feasible, $\DD_H(N\setminus S)\le \dd_H(S)=1$. The only
possible edges inside $H[N\setminus S]$ among original vertices are
precisely the edges of $G[V\setminus C]$. All private leaves in
$R\cup B$ outside $S$ are isolated in $H[N\setminus S]$, because their
centers $r$ and $b$ lie in $S$. Therefore
$\Delta(G[V\setminus C])\le 1$.

Hence $G[V\setminus C]$ is a disjoint union of isolated vertices and
edges. Let $m:=|E(G[V\setminus C])|$. Choose exactly one endpoint from
each of these $m$ edges, and let $X$ be the set of chosen endpoints.
Define $C':=C\cup X$.

Then $C'$ is a vertex cover of $G$. Indeed, every edge of $G$ is either
incident to a vertex of $C$, or else it lies completely inside
$V\setminus C$. The latter edges are precisely the $m$ disjoint edges of
$G[V\setminus C]$, and each of them is hit by $X$.

We now bound $|C'|$. Since $G[V\setminus C]$ has maximum degree at most
$1$, every non-isolated vertex of $G[V\setminus C]$ is an endpoint of
exactly one of those $m$ edges. Therefore
\[
        \left|
        \left\{
        v\in V\setminus C :
        d_{G[V\setminus C]}(v)=1
        \right\}
        \right|
        =
        2m.
\]
Because $\dd_H(S)=1$, the set $T$ is exactly the set of vertices in
$N\setminus S$ whose degree inside $H[N\setminus S]$ equals $1$. The
private leaves outside $S$ have degree $0$, so the vertices counted by
$T$ are precisely the $2m$ endpoints of the edges in $G[V\setminus C]$.
Thus $|T|=2m$.

Let $\ell:=|S\cap(R\cup B)|$ be the number of selected private leaves.
Since $r,b\in S$, we have $|S|=|C|+2+\ell$. Therefore
$|S|+|T|=|C|+2+\ell+2m$. On the other hand, $|C'|=|C|+m$. Hence
\[
        |C'|
        =
        |C|+m
        \le
        |C|+\ell+2m
        =
        |S|+|T|-2.
\]
Thus every feasible solution of value $q\le n+2$ yields a vertex cover
of $G$ of size at most $q-2$.
    
\end{proof}
\medskip

\noindent

\begin{lemma} \label{lem:hard4}
Let $\tau(G)$ be the minimum vertex-cover size of $G$, and let $\OPT(H)$
be the optimum value of the constructed instance $H$.
    $$\OPT(H)=\tau(G)+2.$$
\end{lemma}
\begin{proof}







By \Cref{lem:hard1}, $\OPT(H)\le \tau(G)+2$. Since $\tau(G)\le n$, this implies
$\OPT(H)\le n+2$. Let $S^\star$ be an optimal feasible solution of $H$.
Because $|S^\star|+|T^\star|=\OPT(H)\le n+2$, \Cref{lem:hard3} yields a vertex
cover $C^\star$ of $G$ satisfying
\[
        |C^\star|
        \le
        |S^\star|+|T^\star|-2
        =
        \OPT(H)-2.
\]
Therefore $\tau(G)\le \OPT(H)-2$, or equivalently
$\OPT(H)\ge \tau(G)+2$. Combining both inequalities gives
$\OPT(H)=\tau(G)+2$.
    
\end{proof}
\medskip

\noindent
\paragraph{The reduction}

The construction is polynomial. Moreover, since $G$ has at least one
edge, $\tau(G)\ge 1$. Hence $\OPT(H)=\tau(G)+2\le 3\tau(G)$.

It remains to show that approximate solutions can be converted back with
bounded loss. Let $S$ be any feasible solution of $H$, and let
$q:=|S|+|T|$.

If $q\le n+2$, then by \Cref{lem:hard3} we can construct a vertex cover $C'$ of
$G$ such that $|C'|\le q-2$. Using $\OPT(H)=\tau(G)+2$, we get
\[
        |C'|-\tau(G)
        \le
        q-2-\tau(G)
        =
        q-\OPT(H).
\]

If $q>n+2$, output the trivial vertex cover $C'=V$. Then
$|C'|-\tau(G)=n-\tau(G)$. Since $q>n+2$ and
$\OPT(H)=\tau(G)+2$, we have
\[
        q-\OPT(H)
        >
        n+2-(\tau(G)+2)
        =
        n-\tau(G).
\]
Therefore, in this case as well, $|C'|-\tau(G)\le q-\OPT(H)$.

Thus the reduction is an L-reduction with constants $\alpha=3$ and
$\beta=1$. Since \textsc{Minimum Vertex Cover} is APX-hard, the stated
problem is APX-hard.
\end{proof}

\newpage
\section*{NeurIPS Paper Checklist}

\begin{enumerate}

\item {\bf Claims}
    \item[] Question: Do the main claims made in the abstract and introduction accurately reflect the paper's contributions and scope?
    \item[] Answer: \answerYes{}{} 
    \item[] Justification: See the abstract.

\item {\bf Limitations}
    \item[] Question: Does the paper discuss the limitations of the work performed by the authors?
    \item[] Answer: \answerYes{} 
    \item[] Justification: All assumptions of the paper have been justified in the modelling section (Section~\ref{sec:intro} and~\ref{sec:model}).

\item {\bf Theory assumptions and proofs}
    \item[] Question: For each theoretical result, does the paper provide the full set of assumptions and a complete (and correct) proof?
    \item[] Answer: \answerYes{} 
    \item[] Justification: See~\Cref{app:sec3} and~\Cref{app:sec4}

    \item {\bf Experimental result reproducibility}
    \item[] Question: Does the paper fully disclose all the information needed to reproduce the main experimental results of the paper to the extent that it affects the main claims and/or conclusions of the paper (regardless of whether the code and data are provided or not)?
    \item[] Answer: \answerYes{} 
    \item[] Justification:  \Cref{sec:empirics} gives a detailed description of the experiments, including the dataset and the setup of the data market. 
    We also provide the pseudocode of our algorithm.
    The information is sufficient to reproduce the experimental results.

\item {\bf Open access to data and code}
    \item[] Question: Does the paper provide open access to the data and code, with sufficient instructions to faithfully reproduce the main experimental results, as described in supplemental material?
    \item[] Answer: \answerNo{} 
    \item[] Justification: The California Housing dataset is publicly accessible~\cite{pace1997sparse}, and we have provided sufficient instructions to reproduce the results.
    The paper also includes pseudocode for the implementation.
    However, the current version does not provide code in the supplemental material, and we are planning to release it after acceptance.

\item {\bf Experimental setting/details}
    \item[] Question: Does the paper specify all the training and test details (e.g., data splits, hyperparameters, how they were chosen, type of optimizer) necessary to understand the results?
    \item[] Answer: \answerYes{} 
    \item[] Justification: \Cref{sec:empirics} provides a detailed description of how data is split and assigned to the agents. 


\item {\bf Experiment statistical significance}
    \item[] Question: Does the paper report error bars suitably and correctly defined or other appropriate information about the statistical significance of the experiments?
    \item[] Answer: \answerYes{} 
    \item[] Justification:
     Each experiment is repeated at least 3 times, and results are stable across runs.
     We provide the average result for each instance and report the min and max time across all the instances.

\item {\bf Experiments compute resources}
    \item[] Question: For each experiment, does the paper provide sufficient information on the computer resources (type of compute workers, memory, time of execution) needed to reproduce the experiments?
    \item[] Answer: \answerYes{} 
    \item[] Justification: As stated in~\Cref{sec:empirics}, all experiments are conducted on a MacBook Pro with an Apple M3 chip and 18GB unified memory.
    
\item {\bf Code of ethics}
    \item[] Question: Does the research conducted in the paper conform, in every respect, with the NeurIPS Code of Ethics \url{https://neurips.cc/public/EthicsGuidelines}?
    \item[] Answer: \answerYes{} 
    \item[] Justification: This work is foundational theoretical and empirical research on the data exchange market. 
     It does not involve human subjects, data collection, or any application that would raise specific ethical concerns.

\item {\bf Broader impacts}
    \item[] Question: Does the paper discuss both potential positive societal impacts and negative societal impacts of the work performed?
    \item[] Answer: \answerYes{} 
    \item[] Justification: See~\Cref{sec:intro}
    
\item {\bf Safeguards}
    \item[] Question: Does the paper describe safeguards that have been put in place for responsible release of data or models that have a high risk for misuse (e.g., pre-trained language models, image generators, or scraped datasets)?
    \item[] Answer: \answerNA{} 
    \item[] Justification: The paper does not release any new models, datasets, or other assets that pose a high risk for misuse.

\item {\bf Licenses for existing assets}
    \item[] Question: Are the creators or original owners of assets (e.g., code, data, models), used in the paper, properly credited and are the license and terms of use explicitly mentioned and properly respected?
    \item[] Answer: \answerYes{} 
    \item[] Justification:  The California Housing dataset is publicly accessible~\cite{pace1997sparse} and available for research use.

\item {\bf New assets}
    \item[] Question: Are new assets introduced in the paper well documented and is the documentation provided alongside the assets?
    \item[] Answer: \answerYes{} 
    \item[] Justification: The paper does not introduce new datasets, models, or other assets.

\item {\bf Crowdsourcing and research with human subjects}
    \item[] Question: For crowdsourcing experiments and research with human subjects, does the paper include the full text of instructions given to participants and screenshots, if applicable, as well as details about compensation (if any)? 
    \item[] Answer: \answerNA{} 
    \item[] Justification: The paper does not involve crowdsourcing or research with human subjects.

\item {\bf Institutional review board (IRB) approvals or equivalent for research with human subjects}
    \item[] Question: Does the paper describe potential risks incurred by study participants, whether such risks were disclosed to the subjects, and whether Institutional Review Board (IRB) approvals (or an equivalent approval/review based on the requirements of your country or institution) were obtained?
    \item[] Answer: \answerNA{} 
    \item[] Justification: The paper does not involve human subjects research; no IRB approval is required.

\item {\bf Declaration of LLM usage}
    \item[] Question: Does the paper describe the usage of LLMs if it is an important, original, or non-standard component of the core methods in this research? Note that if the LLM is used only for writing, editing, or formatting purposes and does \emph{not} impact the core methodology, scientific rigor, or originality of the research, declaration is not required.
    \item[] Answer: \answerNA{} 
    \item[] Justification: We used LLM for polishing the pose of text, where the complete draft was fully written by the authors initially and polished later with the help of LLM-based assistants Claude and Gemini.
    We used code assistant Claude Code to implement our original design and ideas.
    All the technical methods, ideas, and core results are entirely the original work of the authors.

\end{enumerate}

\end{document}